\documentclass[twocolumn,trackchanges]{aastex7}

\usepackage{amsmath}
\usepackage{fullpage}
\usepackage{times}
\usepackage{fancyhdr,graphicx,amsmath,amssymb}
\usepackage[ruled,vlined]{algorithm2e}
\usepackage{xfrac}
\usepackage{float}
\include{pythonlisting}

\begin{document}

\title{GraphShed: A Data-derived Graph-based waterShed Group Finder}

\author[orcid=0009-0003-2960-1563,sname='P. Ghafour']{P. Ghafour}
\affiliation{Department of Physics, Shahid Beheshti University,  1983969411, Tehran, Iran}
\email{p.ghafour@outlook.com}
\author[orcid=0000-0002-3655-5417,sname='S. Ansarifard']{S. Ansarifard}
\affiliation{School of Physics, Institute for Research in Fundamental Sciences (IPM), P. O. Box 19395-5531, Tehran, Iran}
\email{ansarifard@ipm.ir}
\author[orcid=0009-0000-4905-7209,sname='M. H. Jalali Kanafi']{M. H. Jalali Kanafi}
\affiliation{Department of Physics, Shahid Beheshti University,  1983969411, Tehran, Iran}
\affiliation{School of Physics, Institute for Research in Fundamental Sciences (IPM), P. O. Box 19395-5531, Tehran, Iran}
\email{jalali@ipm.ir}
\author[orcid=0000-0001-7897-484X,sname='S. M. S. Movahed']{S. M. S. Movahed}
\altaffiliation{Corresponding author}
\affiliation{Department of Physics, Shahid Beheshti University,  1983969411, Tehran, Iran}
\affiliation{School of Astronomy, Institute for Research in Fundamental Sciences (IPM), P. O. Box 19395-5531, Tehran, Iran}
\affiliation{Department of Mathematics and Statistics, The University of Lahore, 1-KM Defence Road, Lahore 54000, Pakistan}
\email[show]{P.Ghafour@outlook.com $ $ $ $ ansarifard@ipm.ir\\jalali@ipm.ir $ $ $ $ m.s.movahed@ipm.ir}

\begin{abstract}
This study introduces GraphShed, a data-derived group-finder that applies top-down watershed segmentation to Voronoi-induced graphs, identifying galaxy systems directly from the density field via a GraphShed-derived linking length, with no density thresholds or tunable parameters. A GraphShed galaxy group catalog compared to a Friends-of-Friends (FoF) catalog built from the IllustrisTNG100-1  simulation with a box size of $\sim 75$ Mpc/h yielding $1067$ and $987$ systems, respectively. The GraphShed-derived linking length adapts to the global clustering power of the sample, varying by $\sim 18 \%$ relative to FoF across different sample selections; thus, catalog differences originate in the watershed segmentation. Cross-catalog comparison shows that $\sim 58 \%$ of FoF systems are identically reproduced by GraphShed, $\sim 5 \%$  are completely removed, and the remainder are partially shaved and/or split into multiple systems. While the $M_{200}$ distributions of the two catalogs are statistically consistent, other structural properties including, $R_{200}$, sphericity, compactness, spin, and centroid shift, differ significantly in at least some richness ranges. The unweighted two-point correlation function of GraphShed systems exhibits a higher amplitude on small scales, $r \lesssim 1$ Mpc/h, and agrees with FoF for $r\gtrsim 5$ Mpc/h.  A velocity-based classification reveals that GraphShed resolves more interacting pairs than FoF, demonstrating its ability to distinguish nearby overdense structures that position-only methods merge into single systems. These results demonstrate that GraphShed preserves cosmological statistics while providing a more resolved detection of galaxy systems and their dynamical interactions.
\end{abstract}

\keywords{\uat{Computational methods}{1965}; \uat{Cosmic web}{330}; \uat{Galaxy groups}{597}; \uat{Galaxy clusters}{584}; \uat{Large-scale structure of the universe}{902}}

\section{Introduction\label{sec:intro}}
Cosmic structure originates from primordial quantum fluctuations \citep{liddle1993cold}, which seeded density perturbations in the early Universe \citep{guth1981inflationary,bernardeau2002large}. Under gravitational instability, these perturbations grow and evolve into the non-linear matter distribution \citep{tomita1967non,shandarin1989large}. Dark matter halos form within this evolving density field, and galaxies, as luminous tracers residing in those halos, provide a biased tracer  to the underlying matter distribution and the cosmological information encoded in it \citep{peebles2020large}.

The spatial distribution of galaxies in the Universe is not uniform. Instead, galaxies are placed in a complex and interconnected network known as the cosmic web \citep{davis1982survey,bond1996filaments}. This intuition has emerged from extensive observational efforts that have mapped the three-dimensional distribution of galaxies with increasing precision. Major surveys, including the 2dFGRS \citep{colless20012df}, SDSS \citep{stoughton2002sloan}, and more recent and ongoing programs such as the DESI \citep{desi2024early}, Euclid \citep{blanchard2020euclid}, and the LSST \citep{marshall2017science}, have provided rich multidimensional datasets that clearly reveal the web-like and hierarchical nature of cosmic structure.

In addition, this cosmic-web is also evident in datasets from large cosmological simulations employing different sets of cosmological parameters inferred from COBE \citep{smoot1992structure}, WMAP \citep{spergel2003first}, and Planck \citep{aghanim2020planck} collaborations. Prominent examples include the Magneticum simulations \citep{bocquet2016halo,dolag2025encyclopedia}, Millennium series (MS: \citealp{springel2005simulations}; MS-II: \citealp{boylan2009resolving}; MS-TNG: \citealp{pakmor2023millenniumtng}) and the Illustris suite (\citealp{nelson2015illustris}; TNG: \citealp{nelson2019illustristng}). These simulations have contributed significantly to identifying and characterizing the large-scale structure by reproducing the emergence of web-like matter distributions within the framework of gravitational structure formation.

These observational and numerical studies consistently demonstrate that the intricate cosmic network is composed of dense knots associated with galaxy groups and clusters, elongated filaments connecting these nodes, and vast under-dense regions known as cosmic voids \citep{libeskind2018tracing,peebles2020large}.

Despite their broad applications, there is no consensus on the formal definition of the cosmic web structures, and different methods exist that apply distinct identification criteria to these datasets (e.g., \citealp{gurzadyan2025cosmic,tavasoli2025filament,peterson2025improving}).

A broad class of methods aims to classify the cosmic web as a whole. Among these, physically motivated approaches rely on dynamical, geometric and topological properties of the underlying density field, which can be constructed using various physical proxies and observables such as mass, luminosity, etc. For instance, tensor-based methods such as the tidal tensor formalism (T-web; \citealp{hahn2007properties}) and the velocity shear tensor formalism (V-web; \citealp{pomarede2017cosmic}) classify regions according to the eigenvalues of the corresponding tensors, typically relative to predefined thresholds, thereby separating the distribution into nodes, filaments, sheets, and voids. Also, topological frameworks quantify the global morphology of the cosmic web using persistent homology and integral geometry. Key descriptors, such as Betti numbers, Euler characteristics, and Minkowski functionals (e.g., \citealp{pranav2019topology}), have been characterized for Gaussian random fields, relating their expected values to critical point statistics and persistence diagrams (e.g., \citealp{pranav2021topology}). In contrast, graph-based approaches construct networks from the galaxy distribution and identify structures using network statistics such as node centrality, connectivity, and related topological features, often combined with threshold criteria on these properties to distinguish different cosmic web morphologies (e.g., \citealp{hong2015network,hong2016discriminating}). Alternative methods include multi-scale morphology-based approaches (e.g., \citealp{aragon2007multiscale,cautun2013nexus}), which classify structures using the hierarchical features of the density field but remain explicitly dependent on smoothing scales and eigenvalue thresholds; and phase-space dynamical approaches (e.g., \citealp{falck2012origami}), which detect shell-crossings and are largely free of density thresholds.

In addition, several methods have been developed to specifically identify individual structures within the cosmic web. For instance, cosmic voids could be detected by approaches that incorporate watershed-based transforms (e.g., \citealp{platen2007cosmic}) and genetic algorithms (e.g., \citealp{ghafour2025vega}), which operate on the density field with multiple parameters. These watershed approaches have also been widely used, including for image segmentation (e.g., \citealp{descombes2017multiple}) and point-process clustering (e.g., \citealp{yang2020individual}). Cosmic filaments can also be extracted using methods, such as topological persistence (\citealp{sousbie2011persistent}), which traces elongated structures in the density field and depends on significance thresholds to filter topological features; or graph-based strategies (e.g., \citealp{ghafour2025gravipast}), which employ techniques such as minimum spanning trees and the local field of the dataset while operating in a parameter-free manner.

Beyond the methods noted earlier, several approaches have been developed to construct group catalogs which have been employed in various studies, ranging from galaxy evolution in dense environments (e.g., \citealp{epinat2024magic}), dwarf satellite statistics (e.g., \citealp{tavasoli2026satellite}), halo assembly bias (e.g., \citealp{croton2007halo}), the calibration of halo occupation statistics (e.g., \citealp{yang2008galaxy}) and testing galaxy-halo connection models (e.g., \citealp{dragomir2018does}), to constraints on cosmological parameters (e.g., \citealp{artis2025srg}). These include graph-based methods such as employing the Minimum Spanning Tree (e.g., \citealp{barrow1985minimal}), where galaxies are connected through a tree that minimizes the total edge length and groups are defined by removing edges longer than a prescribed linking length. Other approaches rely on the concept of shared nearest neighbors (SNN), in which galaxies are associated based on the overlap of their k-nearest neighbor (e.g., \citealp{ertoz2003finding,jarvis2006clustering}), introducing parameters such as the number of neighbors and similarity thresholds. Solely position-based methods, such as the DBSCAN (\citealp{ester1996density}) and its special case, the traditional Friends-of-Friends (FoF; \citealp{davis1985evolution}) algorithm, identify groups through spatial proximity criteria defined by linking lengths or neighborhood radii. These methods are equivalent to single-linkage hierarchical clustering and pruned Minimal Spanning Tree approaches (e.g., \citealp{kaufman2009finding,everitt2011cluster,aggarwal2014data}), which are known to be vulnerable to chaining in the presence of noise. As a result, their ability to capture the full physical complexity of cosmic structures is limited. They might combine nearby over-densities; or connect marginally aligned, filamentary bridges of galaxies, causing these structures to be linked as a single object (\citealp{graham2023group}). This can result in identifying an extended system with a dynamically complex configuration. In this context, irregular clusters are a well-known example that are out of hydrostatic equilibrium and influenced by non‑thermal pressure. Such systems complicate the extraction of reliable physical properties and introduce challenges for both cosmological and astrophysical inference (\citealp{Ansarifard:2019ize}). In addition, the performance of position-based methods remains inherently sensitive to the adopted parameter values. A comprehensive review of group-finder algorithms used in galaxy simulations is provided by \citet{knebe2013structure}.

As discussed above, the majority of existing group-finder methods rely on one or more free parameters or threshold values. These parameters must either be predefined or determined through additional calibration procedures, which can increase computational complexity and resource demands. Although some efforts have been made to adapt parameter choices to the redshift of the dataset (e.g., \citealp{huchra1982groups}), the resulting group catalogs remain sensitive to these selections. Consequently, the dependence of the identified structures on tunable thresholds introduces uncertainties and could affect the robustness of the inferred physical properties.

These limitations highlight the need for an algorithm that operates directly on the density field of the dataset, enabling it to capture the intrinsic features and complexities of the galaxy distribution through its sensitivity to variations in the underlying density field. Such a method should be independent of arbitrary thresholds and free of tunable parameters, minimizing the uncertainties introduced by parameter choices. In addition, it should be straightforward, easy to implement and computationally efficient, enabling its application to large cosmological datasets. An approach with these properties would provide a more robust framework for group identification and could potentially reveal additional physical insights from the resulting group catalogs that remain inaccessible to traditional and parameter-dependent methods.

Motivated by the considerations outlined above, a data-derived, graph‑based watershed approach named GraphShed, is introduced and detailed in Section \ref{sec:GS}. Utilizing this method, the group catalog produced by GraphShed is compared with that obtained from the traditional position‑based FoF approach, in order to probe several aspects of the resulting group catalogs:\\
(I) How the GraphShed-derived linking length is changed compared to that of FoF across samples with various r-band magnitude and stellar mass cuts, using the global clustering power and the effective fraction as diagnostic quantities (Section \ref{sec:res_LL}).\\
(II) How GraphShed alters the assignment of galaxies to groups compared to FoF, assessed using three cross-catalog statistics, including purity, completeness, and removed fraction metrics (Section \ref{sec:res_cc}).\\
(III) How the structural properties of groups detected by the density-field-based GraphShed method compare with those from a position-only approach is assessed through several structural diagnostics of both group catalogs across multiple richness ranges (Section \ref{sec:res_fof}).\\
(IV) How the adoption of GraphShed as the group finder influences large‑scale statistical measures of the group distribution is examined through comparisons of the cumulative mass function and the unweighted two‑point correlation function derived from the two catalogs (Section \ref{sec:LSS}).\\
(V) How the sensitivity of GraphShed to variations in the underlying density field reveals a different population of dynamical interactions including mergers, accretions, and flybys, is determined using a velocity-based classification approach, which is introduced and applied to both catalogs (Section \ref{sec:merger}).\\
The dataset used in this work is described in Section \ref{sec:data}, and the main findings and conclusions are summarized in Section \ref{sec:sum}.

\section{Sample Selection:\\ IllustrisTNG100-1 \label{sec:data}}
The dataset employed in this study is drawn from the gravo-magnetohydrodynamical IllustrisTNG simulation (\citealp{nelson2019illustristng}), which follows the coupled evolution of dark matter, gas, stars, and black holes across cosmic time, from redshift $z=127$ to $z=0$. The simulation is executed using the Arepo moving-mesh framework (\citealp{springel2010pur}) and adopts cosmological parameters consistent with the Planck results (\citealp{ade2016planck}). These parameters include density values of $\Omega_{m}$ = 0.3089 for matter, $\Omega_{b}$ = 0.0486 for baryons, and $\Omega_{\Lambda}$ = 0.6911 for the cosmological constant, along with a scalar spectral index $n_{s} = 0.9667$, the amplitude of matter fluctuations $\sigma_8 = 0.8159$, and a Hubble constant $H = 100 h$ (km/s/Mpc) with $h = 0.6774$.

The analysis is based on the highest-resolution realization within the IllustrisTNG100-1 suite, which spans a co-moving volume cube with a $\sim 75$ (Mpc/h) side length and features the baryonic and dark matter mass resolutions of $1.4\times10^6M_\odot$ and $7.5\times10^6M_\odot$, respectively. This combination of spatial extent and resolution provides a framework for both the robust identification of galaxy groups and clusters and the detailed examination of their diverse characteristics.

From the full IllustrisTNG100-1 galaxy catalog, galaxies are selected based on three criteria: an r-band magnitude brighter than approximately $-16$, a stellar mass exceeding $10^8$ $M_\odot$ and a redshift of $z=0$ (snapshot 99). Applying these thresholds yields a final sample of roughly $41000$ galaxies.

\section{Methodology:\\ GraphShed group finder \label{sec:GS}}

\begin{figure}
\centering
\includegraphics[width=0.48\textwidth]{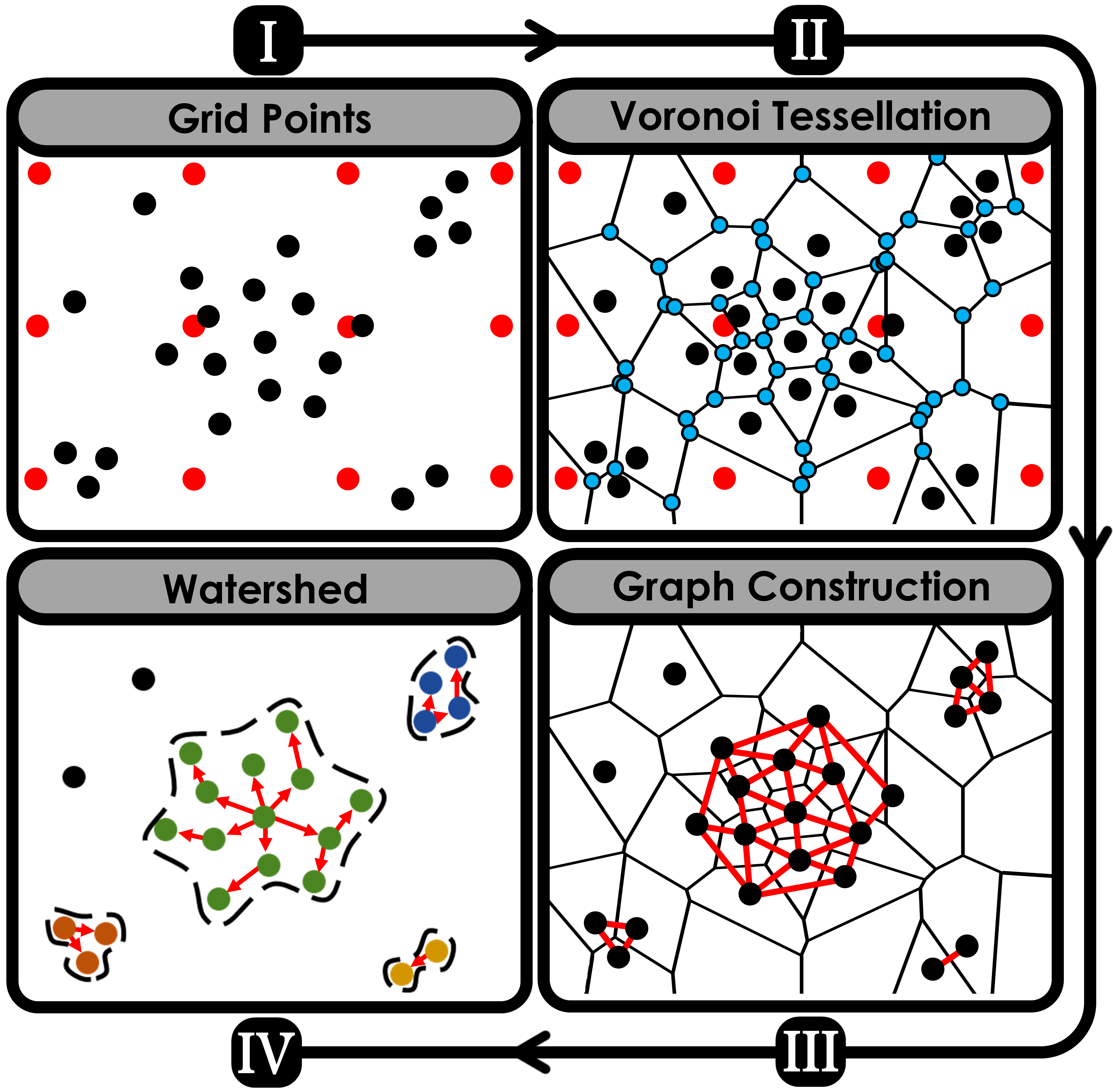}
\caption{Schematic illustration of the main phases involved in the GraphShed group finder. In the first panel ($\mathit{top}$-$\mathit{left}$), grid points with characteristic spacing $\ell$ are added to the galaxy distribution. Grid points and galaxies are shown as red and black dots, respectively. In the second panel ($\mathit{top}$-$\mathit{right}$), the dataset space is partitioned into distinct cells by applying the Voronoi tessellation to the combined set of galaxies and grid points. The vertices of the Voronoi cells are shown as blue dots. In the third panel ($\mathit{bottom}$-$\mathit{right}$), the set of Voronoi-induced graphs are constructed using the shared-cell vertices and the linking length $\mathcal{L}$, with graph edges shown in red. In the final panel ($\mathit{bottom}$-$\mathit{left}$), the Watershed approach is illustrated. Arrows indicate the progression of the algorithm as neighboring galaxies are incorporated into groups. Each identified group is shown in a distinct color and enclosed by dashed lines.}
\label{fig_1}
\end{figure}

In this section, the GraphShed group finder is introduced as a data-derived, easy to implement, and computationally efficient method. It operates on a set of Voronoi-induced graphs and employs a Watershed-based identification process without requiring the adjustment or calibration of free parameters or threshold values. Owing to its structural simplicity, GraphShed can be applied to large galaxy catalogs while requiring reasonable computational demands. The algorithm proceeds through three main phases that are applied sequentially to the input galaxy catalog. These phases are summarized in Algorithm \ref{Met:psudo} and in the schematic diagram shown in Figure \ref{fig_1}, and are described in detail below.

\subsection{Grid points \& Voronoi Tessellation
	\label{sec:Grid_Vor}}

The initial phase of GraphShed consists of four steps. In the first step, a regular cubic grid of auxiliary Steiner points is inserted into the galaxy catalog, as discussed in Appendix \ref{App:Grid}. The characteristic spacing between adjacent grid points along each spatial direction given by $\ell$, which is determined by the mean inter-particle separation, due to the effects of shot noise:
\begin{equation}
	\label{Eq:1}
	\ell \equiv \sqrt[3]{\frac{V_{T}}{n}}
\end{equation}
where $n$ denotes the total number of galaxies in the input catalog and $V_{T}$ represents the total background volume of the dataset.

\begin{algorithm}
	\caption{$GraphShed$ $group$-$finder$}
	\label{Met:psudo}
	\RestyleAlgo{ruled}
	\SetKwSty{textbf}
	\SetArgSty{textnormal}
	\SetNlSty{textbf}{}{}
	\LinesNumbered
	\footnotesize
	\DontPrintSemicolon
	\SetKwComment{Comment}{\# }{}
	\KwIn{$\{x_{i}, y_{i}, z_{i}, m_{i}\}_{i=1}^{N}$: coordinates and stellar masses of $N$ galaxies in the catalog.}
	\KwResult{Group identifier $G_{i}$ for each galaxy $i$.}
	\SetKwData{mono}{mono}
	\Comment{Grid points \& Voronoi tessellation}
	Construct the set of galaxy positions $\mathcal{P}_{i}=(x_{i},y_{i},z_{i})$.\;
	Calculate the characteristic grid spacing $\ell$ by $\mathit{Equation (\ref{Eq:1})}$.\;
	Add a 3D grid with spacing $\ell$ to the input dataset.\;
	Perform Voronoi tessellation on the set of galaxies and grid points $\rightarrow$ Voronoi cells $C_{j}$ for each point $j$ with vertices $X_{j}$.\;
	\ForEach{Voronoi cell $C_{i}$ of galaxy $i$}{
		\If{$\exists$ $\chi$ $\in$ $X_{i}$ such that $\chi$ is outside the dataset boundary \textbf{or} at infinity}{
			Discard $C_{i}$.}
		\Else{$V_{i}$ $\leftarrow$ Volume of $C_{i}$ computed via $\mathit{ConvexHull}$.}}
	\ForEach{Galaxy $i$}{
		Compute the associated mass density by $\mathit{Equation (\ref{Eq:2})}$.}
	\Comment{Graph Construction:}
	Compute the linking length $\mathcal{L}$.\;
	Initialize an un-directed graph $\mathcal{G}=(\mathcal{A},\mathcal{B})$.\;
	\ForEach{Pair of galaxies ($a,b$)}{
		\If{$|X_{a} \cap X_{b}| \geq 2$}{
			Compute distance $D_{ab} \leftarrow \parallel\mathcal{P}_{a}-\mathcal{P}_{b}\parallel$.\;
			\If{$D_{ab} \leq \mathcal{L}$}{
				Add edge $(a,b)$ to graph $\mathcal{G}$.}}}
	\Comment{Watershed Group Identification:}
	Sort galaxies in descending order of their corresponding cell density $\rho_{i}$.\;
	Initialize and mark the group identifier of all galaxies as unassigned: $G_{i} \leftarrow \varnothing$.\;
	\ForEach{Galaxy $k$ in sorted order}{
		\If{$G_{k}$ is already assigned}{
			continue.\;}
		\Else{
			Initialize a new group $\Xi = \{k\}$ with $G_{\Xi}=\xi$.\;
			\Repeat{No new galaxy is added to $\Xi$}{
				\ForEach{Galaxy $q \in \Xi$}{
					$\mathcal{N}(q) \leftarrow \{ f \mid (f,q) \in \mathcal{G} \}$.\;
					\ForEach{Galaxy $f \in \mathcal{N}(q)$}{
						\If{${G}_{f}=\varnothing$}{
							\If{$\rho_{f} \leq \rho_{q}$}{
								Add $f$ to $\Xi$.\;}}}}}
			\ForEach{Galaxy $\nu \in \Xi$}{
				Set $G_{\nu} \leftarrow \xi$.\;}}}
	Return group identifiers $G_{i}$ for all galaxies.
\end{algorithm}

These grids are treated as massless points and therefore do not contribute to the physical density field; however, their inclusion provides several advantages for the tessellation of the dataset space in the subsequent step. It modifies the shape and geometric properties of Voronoi cells by suppressing excessively large cells associated with a data point and reducing irregular, sharply edged cell morphologies. This leads to a more reliable estimation of cell volumes. In addition, it increases the total number of Voronoi cells, thereby enhancing the spatial coverage and accessibility across the dataset space. Also, it modifies boundary conditions by refining the shape of marginal cells near the dataset edges. In particular, it reduces the occurrence of cells associated with galaxies that contain vertices at infinity and mitigates sharply edged peripheral cells, ultimately minimizing the loss of usable volume in boundary regions. These advantages are discussed in detail by \citet{ghafour2025vega}.

In the second step, GraphShed constructs the Voronoi tessellation (\citealp{aurenhammer1991voronoi}) of the combined set of galaxies and inserted grid points. The tessellation is performed in three-dimensional space using the standard Euclidean metric. To accomplish this, the Voronoi diagram of the distribution must be generated. A Voronoi diagram partitions the dataset space into cells surrounding a specified set of objects (in this case, the union of galaxies and grid points). Each point is then associated with a unique Voronoi cell $C_{j}$, defined by its corresponding vertices and encompassing all regions of the dataset space that are closer to that point than to any other. These vertices are shown in Figure \ref{fig_1} as blue dots.

Once the Voronoi cells and their corresponding vertices have been identified, certain marginal cells must be removed (\citealp{ghafour2025vega}). GraphShed first discards all cells containing one or more vertices at infinity, as such cells would possess infinite volume when computing individual cell volumes (step 3). Subsequently, any cell with vertices located outside the dataset (including both galaxies and grid points) boundaries is excluded. These cells often exhibit large or irregular shapes that could impact their computed volumes and, consequently, the resulting density estimates. To prevent the removal of galaxy cells, grid points are placed along the periphery of the galaxy distribution, replicating the same geometry as the galaxy catalog and extending one grid spacing $\ell$ beyond the outermost galaxy positions. This strategy ensures that only the cells associated with grid points are discarded.

The third step focuses on computing the volumes of the cells associated with the galaxies, excluding those corresponding to the inserted grid points. To perform these calculations, GraphShed employs the Quickhull algorithm (\citealp{barber1996quickhull}) to construct the minimum convex polyhedron, or convex hull, surrounding each cell. This construction uses the coordinates of the vertices identified during the Voronoi tessellation in the previous step.

The convex hull of a set of points is defined as the smallest convex polyhedron that fully encloses all points in the set (\citealp{fan1984some}). To obtain the convex hull of a cell, the convex set of its Voronoi-derived vertices is first computed, after which the hull is defined as the minimal convex polyhedron whose boundary contains every vertex of the cell. Once the convex hull is established, its volume is calculated by decomposing the polyhedron into tetrahedrons (\citealp{virtanen2020scipy}). A random interior point is selected, and tetrahedrons are formed by connecting this point to the vertices of each face. Summing the volumes of these tetrahedrons yields the total cell volume $V_{i}$ (\citealp{cohen1979two}). Then, GraphShed computes the mass density\footnote{The density field can be constructed using different physical tracers, including dark matter mass, stellar mass (adopted in this work), luminosity, etc., depending on the dataset used.} associated with each galaxy using the galaxy mass $m_{i}$ and the volume of its corresponding Voronoi cell $V_{i}$, by:
\begin{equation}
	\label{Eq:2}
	\rho_{i} = \frac{m_{i}}{V_{i}}
\end{equation}

\subsection{Linking Length \& Graph Construction
	\label{sec:Graph}}
In the second phase, GraphShed constructs the set of separated graphs. To achieve this, the algorithm first computes the linking length $\mathcal{L}$. This parameter is obtained by calculating the nearest-neighbor (NN) distance for each galaxy $d_{i}^{NN}$ and then taking the mean of only those NN distances that are less than or equal to the mean inter-particle separation $\ell$ (Equation (\ref{Eq:1})), thereby mitigating the influence of galaxies located in highly under-dense regions:
\begin{equation}
	\label{Eq:LL_GS}
	\mathcal{L} \equiv \langle d_{i}^{NN} \rangle, \qquad d_{i}^{NN} \leq \ell
\end{equation}

There are two reasons for adopting this definition instead of the traditional approach, which selects a fraction of the mean inter-particle separation (Equation (\ref{Eq:1}); e.g., \citealp{berlind2006percolation}). First, the traditional method depends solely on the number density of the dataset; thus, datasets with identical number densities but different clustering powers would yield the same linking length. In contrast, the mean nearest-neighbor distance inherently adapts to variations in clustering: among datasets with similar number density, those exhibiting stronger clustering will have shorter nearest-neighbor distances and therefore smaller linking lengths. Second, the traditional approach requires tuning a free parameter, the multiplier applied to the mean inter-particle separation, which is not intrinsically tied to the dataset’s spatial distribution. Different choices of this multiplier can alter the final results, whereas the mean nearest-neighbor distance requires no such adjustable parameters.

The GraphShed uses the Voronoi cells corresponding to the galaxies $C_{i}$ to identify each galaxy’s neighbors, in the final step of second phase. The algorithm examines the vertices of the Voronoi cells, and whenever two cells share two or more common vertices ($|X_{a} \cap X_{b}| \geq 2$), the galaxies associated with those cells are considered neighbors. This includes each cell's face-neighbors, defined as cells sharing a face and thus having three or more common vertices, and its edge-neighbors defined as cells sharing a side and therefore having exactly two common vertices (\citealp{berlyand2005discrete}). If the distance between any two neighboring galaxies is less than or equal to the linking length ($D_{ab} \leq \mathcal{L}$), GraphShed adds an edge between that pair of galaxies to the set of separated graphs.

\subsection{Watershed Group Identification
	\label{sec:Wat}}
In the final phase, GraphShed applies the Watershed method (\citealp{beucher1979use,dougherty1992mathematical}) to the set of separated graphs to identify groups of galaxies. The procedure begins with the galaxy that possesses the highest cell density in the dataset, the global maximum. Using the graph constructed in the previous phase, the algorithm identifies its neighboring galaxies and adds those with densities less than or equal to that of the current galaxy, following a top-down progression. This approach enables groups to grow outward from density maxima, thereby differentiating distinct nearby peaks in a manner similar to some physical density-based group and subhalo finders (e.g., \citealp{springel2001populating,graham2023group}), rather than spuriously merging adjacent structures, as occurs in methods that operate solely based on the positions of galaxies (e.g., \citealp{ester1996density}). This process is repeated for each newly added galaxy and continues until no additional galaxies satisfy the density condition or remain connected to the growing structure.

Once the identification process of a group is complete, the algorithm proceeds to the next unassigned global maximum in the density field and applies the same procedure to identify another group. This continues until no galaxies remain that possess at least one connected edge and therefore the potential to form a distinct group.

\section{Results and Discussion\label{sec:res}}
This section presents the resulting group catalog obtained from applying the GraphShed group finder, as detailed in Section \ref{sec:GS}, to the dataset described in Section \ref{sec:data}.  For comparison, a group catalog is also constructed using the Friends-of-Friends (FoF) algorithm (\citealp{davis1985evolution}). The FoF method is selected for comparison due to its widespread use in constructing galaxy group and cluster catalogs in previous studies (e.g., \citealp{pizzardo2023illustristng,levis2025galaxy,masson2026calibrating}). FoF’s reliance on a position-only approach, makes it well suited for highlighting the differences between catalogs identified using a physical density-based method and those obtained from purely positional algorithms. In contrast to GraphShed, the FoF method does not incorporate any physical field, relies solely on galaxy positions and identifies groups using a linking length metric.

Initially, the GraphShed-derived linking length $\mathcal{L}$ (Equation (\ref{Eq:LL_GS})) is compared with that of FoF (Section \ref{sec:res_LL}) using various samples with different r-band absolute magnitude and stellar mass cuts. Afterward, the cross-catalog statistics between the group catalogs of the two methods are quantified (Section \ref{sec:res_cc}) using the completeness, purity, and removed fraction metrics. Subsequently, structural properties of the groups identified by GraphShed and FoF are compared (Section \ref{sec:res_fof}), including $R_{200}$, $M_{200}$, sphericity, compactness, spin parameter and centroid shift. Then, the large-scale statistical properties of the resulting catalogs are analyzed at Section \ref{sec:LSS}. In particular, the halo mass functions and two-point correlation functions derived from the GraphShed and FoF catalogs are compared. This analysis allows an assessment of whether differences in the internal structure of the groups and clusters identified by the two methods propagate to statistical measures describing the large-scale distribution of matter. Finally, Section \ref{sec:merger} presents a comparison of mergers, flybys, and accretion events detected in the two catalogs. These interactions are determined using a velocity‑based merger identification procedure, introduced and applied to the GraphShed and FoF group catalogs; enabling a direct comparison of the resulting merger classifications obtained from the two methods. 

A summary of the metrics used to compare GraphShed and FoF catalogs across different aspects is provided in Table \ref{tab:0}, with the associated details discussed in the following subsections.

\begin{deluxetable}{c|c}
	\tablecaption{Types and metrics used to compare GraphShed and FoF catalogs in this study. See the text for more details. \label{tab:0}}
	\tablewidth{0pt}
	\tablehead{
		\colhead{Type} &  \colhead{Metrics}
		}
	\startdata
	Clustering & $\{\delta_{CE}, \delta_{b}\}$ \\
	Cross-Catalog & $\{P_{\alpha}, C_{\alpha}, R_{\alpha}\}$ \\
	Structural & $\{R_{200}, M_{200}, \Theta, \zeta, \lambda^{\prime}, \Delta\mathcal{C}\}$ \\
	Large-Scale & $\{\xi(r), \mathcal{N}(>M_{200}) \}$
	\enddata
\end{deluxetable}

\subsection{Linking Lengths
	\label{sec:res_LL}}
As discussed in Section \ref{sec:Graph}, the GraphShed-derived linking length is calculated using Equation (\ref{Eq:LL_GS}), while the FoF linking length is typically defined as a constant fraction $b$ of the mean inter-particle separation $\ell$ (Equation (\ref{Eq:1})). In this study, the $b=0.2$ is adopted, consistent with previous studies incorporating galaxy catalogs at redshift $z=0$ (e.g., \citealp{more2011overdensity,yoo2025tracing}). For the comparison of the two methods' linking lengths, two quantities are introduced and measured: the global clustering power $\delta_{CE}$, and the percentage change of the GraphShed-derived linking length with respect to that of FoF $\delta_{b}$ (Table \ref{tab:0}).

The $\delta_{CE}$ is defined based on the 3-dimensional extension of the Clark-Evans index (\citealp{clark1954distance,clark1979generalization}), introduced as the comparison of the mean nearest-neighbor distance of all the galaxies inside a sample $\langle d_{S}^{NN} \rangle$, to that expected from a completely random and homogeneous Poisson distribution, given by the following equation as detailed in Appendix \ref{App:CE}:
\begin{equation}
	\label{d_CE}
	\langle d_{P}^{NN} \rangle = \Gamma \left( \frac{4}{3} \right) \left(\frac{3V_{T}}{4\pi N_{T}}\right)^{1/3}
\end{equation}
where $\Gamma$ is the gamma function, and  $V_{T}$ and $N_{T}$ are the total volume and total number of galaxies inside the sample. The global clustering power is then defined as:
\begin{equation}
	\label{del_CE}
	\delta_{CE} \equiv 1 - \frac{\langle d_{S}^{NN} \rangle}{\langle d_{P}^{NN} \rangle}
\end{equation}
thus, if a sample is highly clustered, it will have a lower $\langle d_{S}^{NN} \rangle$ and therefore a higher $\delta_{CE}$, and vice versa. For the calculation of $\delta_{b}$, the effective fraction $b_{\mathcal{L}} \equiv \frac{\mathcal{L}}{\ell}$ is first defined, where $\mathcal{L}$ is the GraphShed-derived linking length (Equation (\ref{Eq:LL_GS})) and $\ell$ is the mean inter-particle separation (Equation (\ref{Eq:1})). Then $\delta_{b}$ is defined as the percentage change of the $b_{\mathcal{L}}$ from the FoF's traditional $b=0.2$ value:
\begin{equation}
	\label{del_b}
	\delta_{b} \equiv \left(1 - \frac{b_{\mathcal{L}}}{0.2}\right) \times 100 .
\end{equation}

In addition, this comparison is performed using various samples from the TNG100-1 simulation (Section \ref{sec:data}), with different r-band absolute magnitude ($M_{r}$) cuts ranging from $-14$ to $-20$, and stellar mass ($M_{*}$) cuts ranging from $10^{7} M_{\odot}$ to $10^{10} M_{\odot}$.

\begin{figure}
	\centering
	\includegraphics[width=0.47\textwidth]{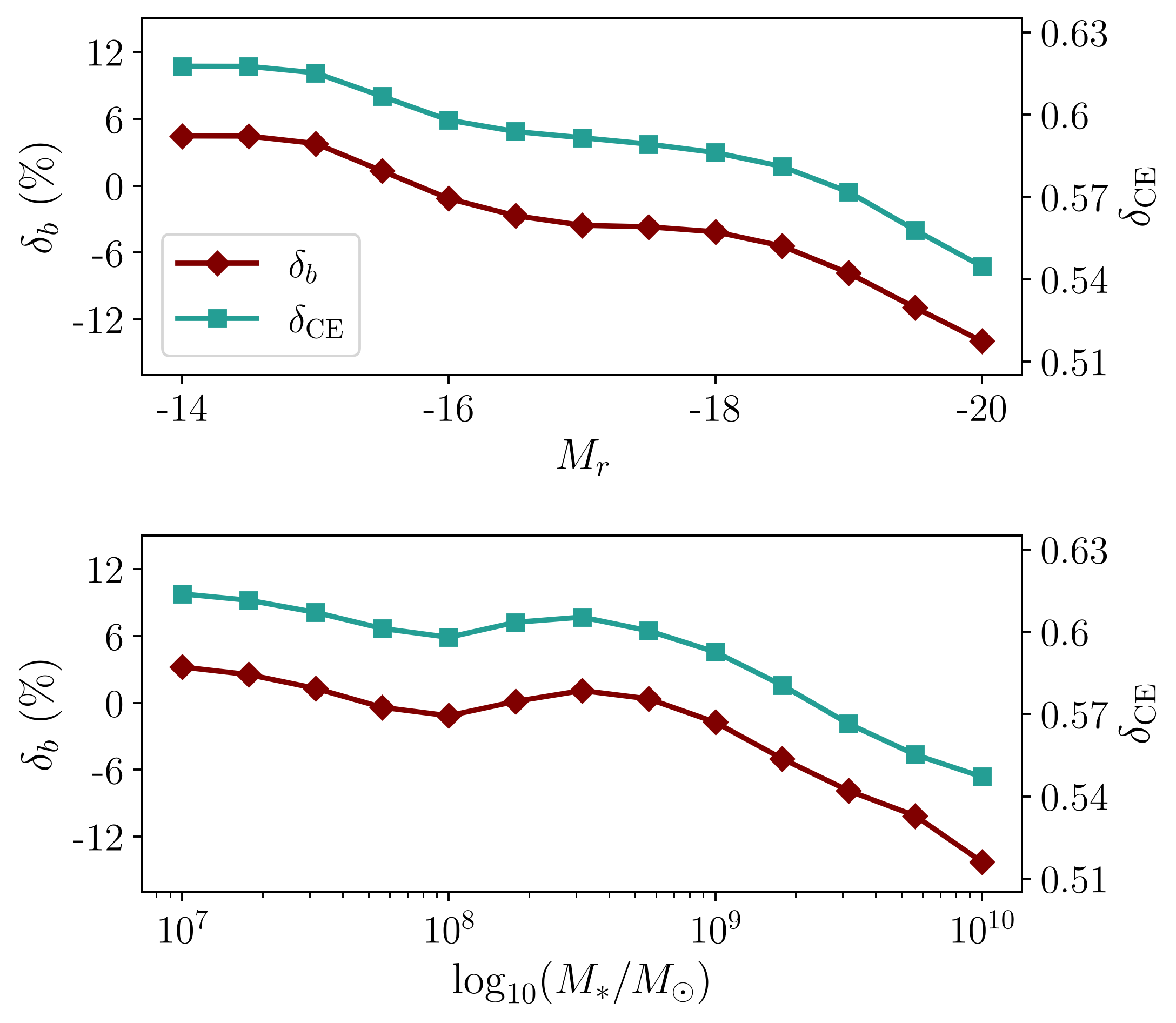}
	\caption{Comparison of $\delta_{b}$ (left y-axis, Equation (\ref{del_b})) and $\delta_{CE}$ (right y-axis, Equation (\ref{del_CE})) for different sample selections. Variation with r-band absolute magnitude cut is presented at $top$ panel, while keeping the stellar mass cut fixed at $M_{*}=10^{8}M_{\odot}$. Variation with stellar mass cut is illustrated at the $bottom$ panel, while keeping the r-band magnitude cut fixed at $M_{r}=-16$.}
	\label{fig_LL}
\end{figure}

The outcomes of measuring these quantities on the samples mentioned above are illustrated in Figure \ref{fig_LL}. First, a fixed stellar mass cut of $10^{8} M_{\odot}$ is considered and only the r-band absolute magnitude cut is varied, as shown in the top panel of this figure. The $\delta_{CE}$ decreases from about $0.62$ at $M_{r}=-14$ to about $0.54$ at $M_{r}=-16$, indicating that the global clustering power is reduced when brighter magnitude cuts are applied. The $\delta_{b}$ correspondingly decreases from about $4.4\%$ to about $-13.9\%$ over the same $M_{r}$ range. Then, a fixed r-band magnitude cut of $-16$ is considered while the stellar mass cut is varied, as presented in the bottom panel of Figure \ref{fig_LL}. Increasing the stellar mass cut toward higher mass thresholds results in a decrease in both $\delta_{CE}$ and $\delta_{b}$, from about $0.61$ and $3.2\%$ at $M_{*}=10^{7}M_{\odot}$ to about $0.55$ and $-14.3\%$ at $M_{*}=10^{10}M_{\odot}$, respectively.

These results show that the GraphShed-derived linking length adapts to the global clustering power of the input sample (as mentioned in Section \ref{sec:Graph}), adjusting its value accordingly. Furthermore, they demonstrate that the effective fraction $b_{\mathcal{L}}$, remains close to the FoF's traditional value of $0.2$ but varies by approximately $18.3\%$ and $17.5\%$ between the highest and lowest clustered samples analyzed by varying the r-band magnitude and stellar mass cuts, respectively. Lower $b_{\mathcal{L}}$ values correspond to higher clustering power.

For the rest of the comparison, dataset mentioned in Section \ref{sec:data}, has a $\delta_{b} \approx -1\%$. The GraphShed-derived grid spacing, linking length and the FoF linking lengths for this dataset are calculated as $\sim2.16$, $\sim0.437$ and $\sim0.433$ Mpc/h, respectively. These close values ensure that the comparisons are largely insensitive to effects arising from differences in the adopted linking lengths of the two methods.

\subsection{Cross-Catalog Statistics 
	\label{sec:res_cc}}
	
For a consistent comparison between the catalogs produced by the GraphShed and FoF methods, those groups containing more than or equal to $5$ members are considered. This restriction is adopted to avoid potential misinterpretations arising from very small groups, since their properties cannot be reliably measured (e.g., \citealp{seppi2025modelling}). Such a minimum group richness threshold is commonly adopted in cosmological and astrophysical studies (e.g., \citealp{fernandez2024revealing,marini2025detecting}). With this criterion, the number of detected groups is $1067$ for GraphShed and $987$ for the FoF method. The group richness ($N$) in these catalogs spans from small systems with $5$ members to massive galaxy clusters containing up to $211$ and $402$ galaxies for GraphShed and FoF, respectively. The difference in the maximum richness captured by the two methods arises from the position-only approach of the FoF algorithm, which might combine multiple nearby over-densities into a single group (\citealp{graham2023group}). In contrast, the sensitivity of the GraphShed method to variations in the density field of the dataset, combined with its top-down watershed segmentation approach, could separate such over-densities into distinct systems.

To quantify the correspondence between galaxy memberships identified by GraphShed and FoF, the two group catalogs are compared using a set of cross-catalog statistics (Table \ref{tab:0}). These metrics characterize how the galaxy members assigned to a group in one catalog are distributed among groups in the other, thereby providing quantitative measures of membership agreement and differences between the two group-finding methods. In particular, the purity and completeness metrics are adopted from previous studies (e.g., \citealp{looser2021optimizing,ma2025widely}), while the removed fraction is introduced in this work to complement the analysis. These metrics are defined and measured as follows. In the definitions below, $\alpha$ and $\beta$ denote the two catalogs being compared. Throughout this subsection, metrics pertaining to the FoF and GraphShed catalogs are denoted with subscripts $FoF$ and $GS$, respectively.

The purity $P_{\alpha}$ is defined as the fraction of galaxies inside each $\alpha$ group that belong to its dominant counterpart in the $\beta$ catalog. This metric reveals that while $P_{FoF}$ ranges from $0.32$ to $1$, $P_{GS}$ is equal to $1$ for all GraphShed groups.

The completeness $C_{\alpha}$ is defined as how completely each group in the $\alpha$ catalog recovered the galaxies in its dominant counterpart group from the $\beta$ catalog. This metric shows that while $C_{GS}$ ranges from $0.11$ to $1$, $C_{FoF}$ is equal to $1$ for all FoF groups.

Finally, the removed fraction $R_{\alpha}$ is defined as the fraction of galaxies in each $\alpha$ group that are not assigned to any group in the $\beta$ catalog. This metric indicates that while $R_{FoF}$ ranges from 0 to 1, the $R_{GS}$ is consistently $0$.

\begin{figure}
	\centering
	\includegraphics[width=0.43\textwidth]{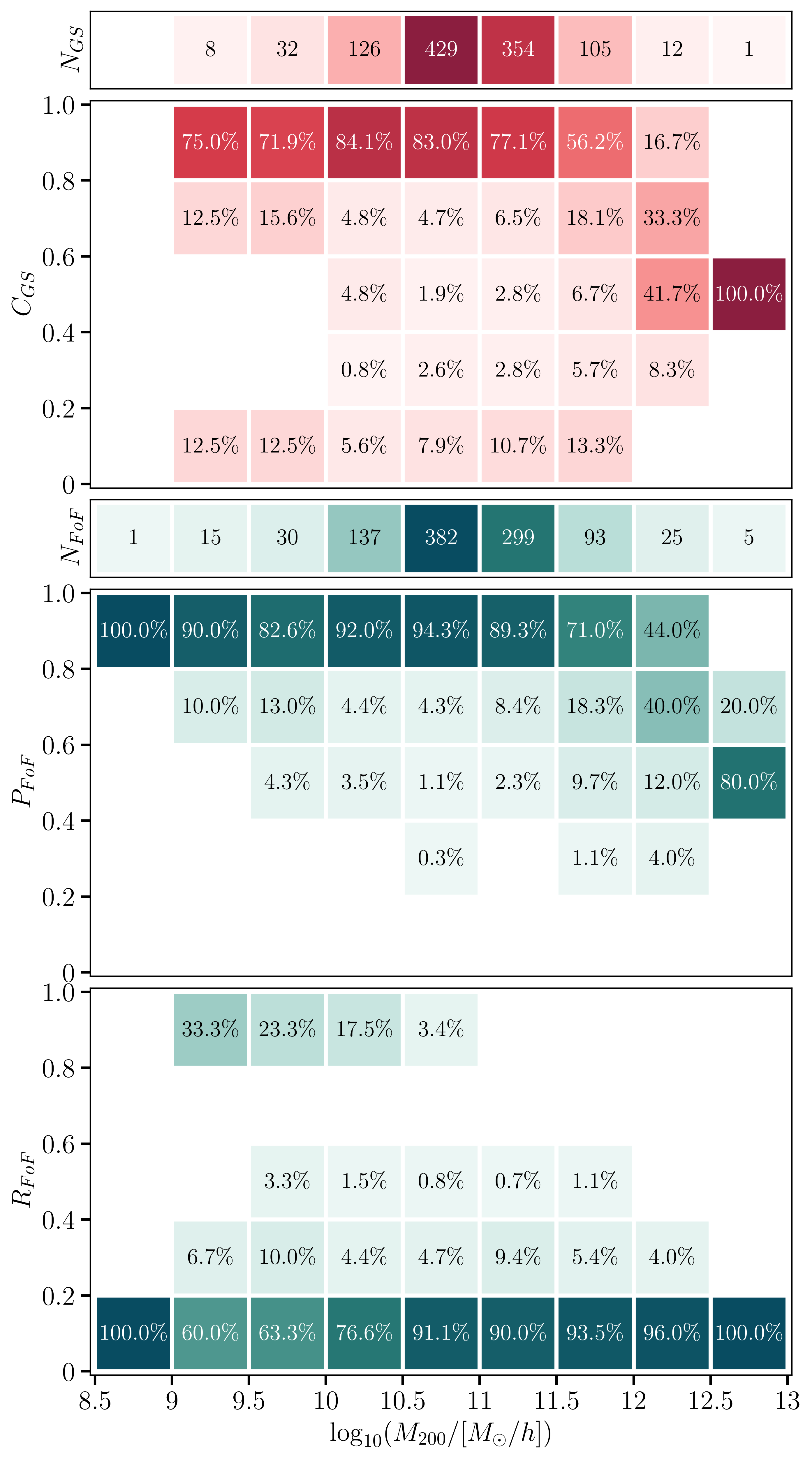}
	\caption{Distribution of the cross-catalog membership statistics as a function of group mass. The $upper$, $middle$, and $bottom$ panels show the GraphShed completeness $C_{GS}$, the FoF purity $P_{FoF}$ and removed fraction $R_{FoF}$, respectively. In each mass bin, the percentages represent the fraction of groups falling within the corresponding metric intervals, normalized by the total number of groups in that mass bin. The total number of groups in each mass bin is indicated by $N_{GS}$ and $N_{FoF}$ panels for the GraphShed and FoF catalogs, respectively.}
	\label{fig_cc}
\end{figure}

Mentioned ranges for the metrics, demonstrate that while some groups are identically recovered by both methods, with both purity and completeness equal to $1$ and constituting approximately $58 \%$ of the FoF catalog, there exist groups that are completely removed by GraphShed (about $5 \%$), as well as cases where GraphShed shaves the low-density outskirts off the FoF systems and segments what remains.

The results are presented in of Figure \ref{fig_cc}, which show the GraphShed groups completeness $C_{GS}$ (upper panel), and the FoF groups purity $P_{FoF}$ (middle panel) and removed fraction $R_{FoF}$ (bottom panel), binned by the group $M_{200}$ mass. Other metrics are not shown due to their constant value for all groups. Each $M_{200}$ bin is normalized by the total number of groups within it, while the total number of groups in each bin is indicated in the $N_{GS}$ and $N_{FoF}$ panels. As can be seen in this figure, the high percentages in the highest bin of $C_{GS}$ and $P_{FoF}$, and in the lowest bin of $R_{FoF}$, are driven by the presence of identical systems identified by both methods. Conversely, the highest bin of $R_{FoF}$ corresponds to groups that are completely removed by GraphShed. Furthermore, the wide distribution range of $C_{GS}$ across almost all $M_{200}$ bins indicates that GraphShed segments FoF systems into multiple nearby over-densities that are in close positional proximity. While FoF considers these as a single system due to its insensitivity to variations in the density field, GraphShed is sensitive to such variations and thus splits them accordingly.

An additional point of Figure \ref{fig_cc} is the dependence of these metrics on group mass. At low and intermediate masses, most groups exhibit high membership correspondence, with the majority of GraphShed and FoF groups having $C_{GS} \geq 0.6$ and $P_{FoF} \geq 0.6$, respectively. This correspondence becomes progressively broader toward higher $M_{200}$ bins, with the fraction of groups satisfying these criteria decreases at $M_{200} \geq 10^{11.5} M_{\odot}/h$ bins. The $R_{FoF}$ distribution shows a different trend, which is dominated by low $R_{FoF}$ values at intermediate and high $M_{200}$ bins, while this dominance is weaker at low $M_{200}$ bins. Thus, while the two catalogs exhibit generally higher agreement at low and intermediate $M_{200}$ bins, differences in their membership assignments become more evident toward higher $M_{200}$ bins, indicating that the strongest discrepancies between the two catalogs are preferentially associated with massive systems.

\subsection{Structural Properties
	\label{sec:res_fof}}
The structural comparison here includes the basic group properties of radius $R_{200}$ and mass $M_{200}$, along with the shape and dynamical metrics sphericity, compactness, spin parameter, and centroid shift (Table \ref{tab:0}). Since all of these quantities are estimated from the member galaxies assigned to each group, their measurements can be affected by group richness. Moreover, as mentioned in Section \ref{sec:res_cc}, the two catalogs span different ranges of group richness, which can introduce a selection effect when their structural properties are compared. To minimize this effect and ensure a more consistent comparison, the results are presented in four different richness bins (e.g., \citealp{li2019effect}) $5 \leq N_{G} \leq 10$, $10 < N_{G} \leq 20$, $20 < N_{G} \leq 50$ and $50 < N_{G}$ as noted in Table \ref{tab:1}. These four bins contain $688$, $224$, $117$ and $38$ GraphShed groups, and $616$, $208$, $112$ and $51$ FoF systems, respectively.

Each of structural properties is probed across all four richness bins using the two-sided Kolmogorov--Smirnov test (hereafter KS; \citealp{massey1951kolmogorov}). The KS test provides a $p$-value for assessing statistical significance, while the $D$-statistic represents the maximum vertical distance between the cumulative distribution functions of the two group catalogs, providing a measure of effect size and the magnitude of their distributional difference (\citealp{wall2012practical}). For a robust evaluation of these metrics and to determine whether the two distributions are significantly different, a significance threshold of $\alpha$ is adopted for the $p$-values. The corresponding threshold for the D-statistics is then calculated by (e.g., \citealp{husser2020stellar}):
\begin{equation}
	\label{Eq:d_cr}
	D_{\rm crit} = c(\alpha) \sqrt{\frac{n_{\rm GS}+n_{\rm FoF}}{n_{\rm GS}n_{\rm FoF}}},
\end{equation}
where $c(\alpha)=1.36$ for $\alpha=0.05$ \citep{gehan2021automated}, and $n_{\rm GS}$ and $n_{\rm FoF}$ are the total number of systems in the GraphShed and FoF catalogs within each richness bin, as mentioned earlier. This yields $D_{\rm crit}$ values of $0.08$, $0.13$, $0.18$, and $0.29$ for the four richness bins in increasing order of $N_G$. The resulting $p$-values, $D$-statistics, and corresponding $D_{\rm crit}$ values, are presented in Table \ref{tab:1}.

\begin{figure}
	\centering
	\includegraphics[width=0.47\textwidth]{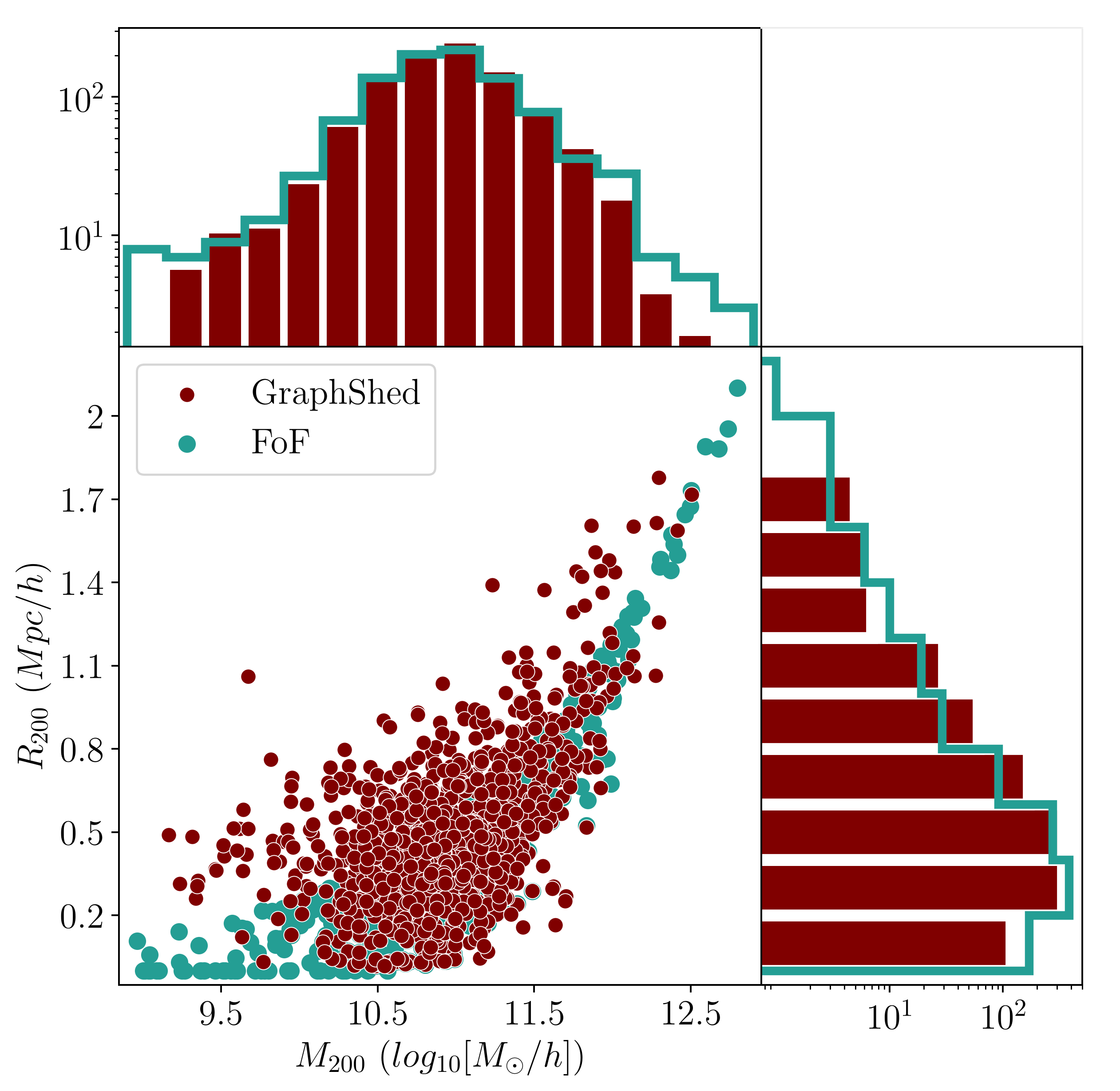}
	\caption{The main panel illustrates the distribution of the $R_{200}$ radius with respect to the stellar mass $M_{200}$ for groups identified by the GraphShed and FoF methods, in dark red and light blue, respectively. The corresponding logarithmic histograms of the $M_{200}$ and $R_{200}$ distributions are presented using the same color scheme in the upper and right panels, respectively.}
	\label{fig_2}
\end{figure}

The comparison begins with the $R_{200}$ radii and $M_{200}$ masses, as these serve as the basic group properties for other metrics. The $R_{200}$ is defined as the radius from the Center-Of-Mass (COM) of a group within which the stellar mass density is approximately $200$ times the background. Accordingly, $M_{200}$ corresponds to the total stellar mass of galaxies contained within the $R_{200}$ radius. The overall distribution of these parameters for the resulting galaxy group catalogs of GraphShed and FoF are shown in Figure \ref{fig_2}. The $R_{200}$ radii of the GraphShed groups extend up to approximately $1.78$ Mpc/h, whereas those of the FoF groups reach about $2.11$ Mpc/h. Similarly, the $M_{200}$ mass range of the GraphShed and FoF groups extends to approximately $10^{12.51}$ and $10^{12.79}$ $M_{\odot}/h$, respectively. These results suggest that FoF constructs some relatively large and massive clusters, while GraphShed segments multiple over-dense regions that FoF groups together (as discussed in Section \ref{sec:res_cc}). Consequently, this leads to reduced upper bounds in the $R_{200}$ radius and $M_{200}$ mass distributions of the GraphShed-identified galaxy groups and clusters. The corresponding KS tests across the four richness bins yield the $p$‑values and $D$-statistics listed in the first two rows of Table \ref{tab:1}. For the two lowest richness bins, the obtained $p$-values are smaller than the adopted significance level $\alpha=0.05$, while the corresponding $D$-statistics exceed their respective $D_{\rm crit}$ values, indicating a statistically significant difference between the two catalogs. In contrast, for the two highest richness bins, the $p$-values exceed $\alpha$ and the corresponding $D$-statistics remain below $D_{\rm crit}$, indicating no statistically significant difference between the $R_{200}$ distributions. For the $M_{200}$, the resulting $p$-values and $D$-statistics imply that the mass distributions of the groups identified by GraphShed and FoF are statistically consistent across all $N_{G}$ bins.

\begin{deluxetable}{c|cc|cc|cc|cc}
	\tablecaption{Two-sided KS-test metrics, p-value, D-statistics and $D_{crit}$ are presented for structural parameters of GraphShed and FoF galaxy systems, binned by group richness $N_{G}$.\label{tab:1}}
	\tablewidth{0pt}
	\tablehead{
		\colhead{$N_{G}$} & \multicolumn{2}{|c}{$5 \leq N_{G} \leq 10$} & \multicolumn{2}{|c}{$10 < N_{G} \leq 20$} & \multicolumn{2}{|c}{$20 < N_{G} \leq 50$} & \multicolumn{2}{|c}{$50 < N_{G}$}\\
		\hline
		\colhead{$D_{\rm crit}$} & \multicolumn{2}{|c}{$0.08$} & \multicolumn{2}{|c}{$0.13$} & \multicolumn{2}{|c}{$0.18$} & \multicolumn{2}{|c}{$0.29$}\\
		\hline
		\colhead{Parameter} & \multicolumn{1}{|c}{p-value} & \multicolumn{1}{c}{D-Stat.} & \multicolumn{1}{|c}{p-value} & \multicolumn{1}{c}{D-Stat.} & \multicolumn{1}{|c}{p-value} & \multicolumn{1}{c}{D-Stat.} & \multicolumn{1}{|c}{p-value} & \multicolumn{1}{c}{D-Stat.}
	}
	\startdata
	$\mathrm{R}_{200}$ & $1.55 \times 10^{-13}$ & $0.21$ & $2.04 \times 10^{-4}$ & $0.18$ & $0.07$ & $0.16$ & $0.93$ & $0.11$\\
	$\mathrm{M}_{200}$ & $0.13$ & $0.06$ & $0.63$ & $0.07$ & $0.89$ & $0.08$ & $0.11$ & $0.24$\\
	$\mathrm{Sphericity}$ & $5.62 \times 10^{-3}$ & $0.09$ & $8.71 \times 10^{-5}$ & $0.21$ & $1.34 \times 10^{-5}$ & $0.32$ & $1.33 \times 10^{-5}$ & $0.49$\\
	$\mathrm{Compactness}$ & $3.72 \times 10^{-14}$ & $0.22$ & $1.06 \times 10^{-5}$ & $0.23$ & $8.34 \times 10^{-6}$ & $0.32$ & $4.45 \times 10^{-5}$ & $0.47$\\
	$\mathrm{Spin}$ & $3.16 \times 10^{-4}$ & $0.11$ & $0.21$ & $0.05$ & $0.01$ & $0.19$ & $2.9 \times 10^{-4}$ & $0.43$\\
	$\mathrm{Centroid}$ $\mathrm{Shift}$ & $0.93$ & $0.03$ & $0.19$ & $0.09$ & $0.03$ & $0.19$ & $0.02$ & $0.31$\\
	\enddata
\end{deluxetable}

\begin{figure*}
	\centering
	\includegraphics[width=0.999\textwidth]{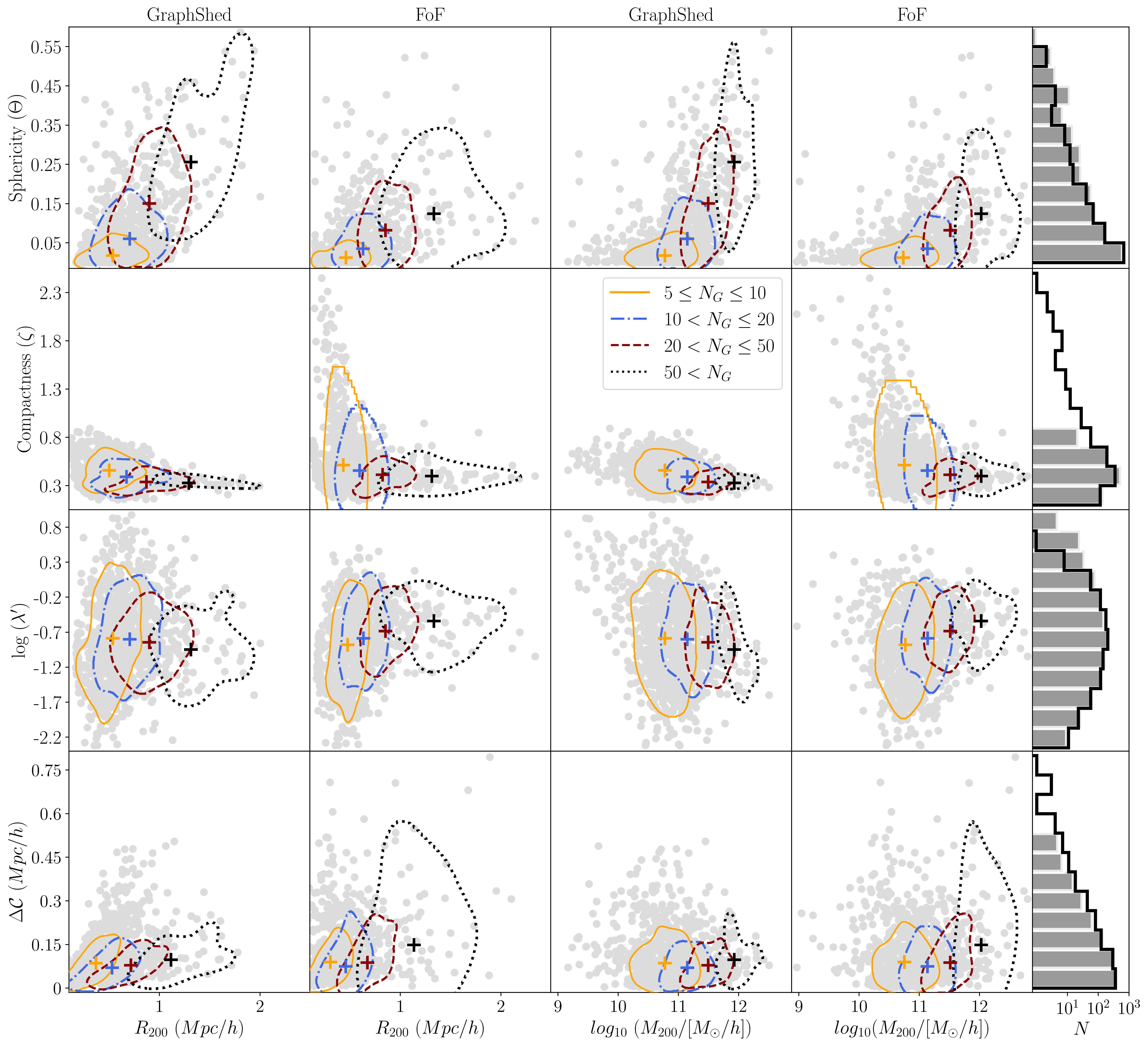}
	\caption{The distributions of sphericity ($\Theta$; $Upper$ row), compactness ($\zeta$; $Second$ row), spin parameter ($\lambda^{\prime}$; $Third$ row) and centroid shift ($\Delta \mathcal{C}$; $Bottom$ row) for groups identified by GraphShed and FoF, shown as a function of $R_{200}$ radii ($First$ and $Second$ columns) and $M_{200}$ stellar mass ($Third$ and $Fourth$ columns), respectively. In all panels, the full distribution for each catalog (indicated above each column) is shown as grey background scatter points, while the distributions for groups in each richness bin are represented by their median ($+$ sign) and $1\sigma$ contours in different colors and line styles. The logarithmic histograms of the total distribution for each parameter are shown in the rightmost column of each row, with GraphShed and FoF shown in dark grey and black, respectively.}
	\label{fig_3}
\end{figure*}

For the comparison of the shapes of the resulting galaxy groups and clusters, the sphericity ($\Theta$; Equation (\ref{Eq:3})) and compactness ($\zeta$; Equation (\ref{Eq:4})) metrics are measured. The sphericity is defined as:
\begin{equation}
\label{Eq:3}
\Theta \equiv \frac{c}{a}
\end{equation}
where $a$ and $c$ denote the largest and smallest eigenvalues of the group’s moment of inertia tensor (\citealp{hahn2007properties}). Higher values of $\Theta$ correspond to more spherical group morphologies. The overall distribution of the sphericity values of galaxy groups in the GraphShed and FoF catalogs and are shown in the top row of Figure \ref{fig_3} with respect to the groups’ $R_{200}$ radius at first two columns and $M_{200}$ mass at third and fourth columns, respectively. In addition, the median and $1\sigma$ contours for different richness bins are shown using distinct colors and line-styles. As illustrated in these panels, the sphericity of the groups generally increases with increasing $R_{200}$, $M_{200}$, and richness in both catalogs. This trend becomes more pronounced with increasing $R_{200}$, $M_{200}$, and $N_G$ for the GraphShed groups, with the difference between the two catalogs becoming steadily larger and more evident. Notably, the FoF catalog contains several groups with intermediate-to-large $R_{200}$, $M_{200}$, and $N_G$ values that exhibit relatively low sphericity, a feature that is not observed in the GraphShed group catalog.The KS tests applied to the $\Theta$ distributions across the four $N_G$ bins yield the $p$-values and $D$-statistics presented in the third row of Table \ref{tab:1}. Across all four bins, these metrics indicate statistically significant differences between the sphericity distributions of the galaxy groups identified by GraphShed and FoF. Moreover, the $D$-statistic increases from the lowest- to the highest-richness bin, indicating that the magnitude of the distributional difference between the two catalogs becomes progressively larger with increasing group richness, consistent with the trends illustrated in Figure \ref{fig_3}.

The compactness parameter is then defined as the mean radial distance of the group members normalized to the $R_{200}$:
\begin{equation}
\label{Eq:4}
\zeta \equiv \frac{1}{R_{200}\times N} \sum_{i=1}^{N} ||\vec{r}_{i} - \vec{r}_{COM}||
\end{equation}
where $\vec{r}_{i}$ and $\vec{r}_{COM}$ represent the position vectors of the group members and the center of mass, respectively. Lower values of $\zeta$ indicate that the member galaxies are located closer to the COM, corresponding to a more compact system. The resulting compactness of the groups identified by GraphShed and FoF is illustrated in the second row of Figure \ref{fig_3} as a function of the groups’ $R_{200}$ and $M_{200}$, following the same configuration as above, with the medians and $1\sigma$ contours for the different $N_G$ bins shown. As can be seen from these panels, the groups identified by GraphShed generally exhibit lower $\zeta$ values and, consequently higher compactness compared to those identified by FoF. Overall, a decreasing trend in the $\zeta$ parameter is evident with increasing $R_{200}$, $M_{200}$, and $N_G$, indicating that larger and richer systems tend to be more compact than smaller and less-rich groups in both catalogs. This trend becomes weaker toward larger $R_{200}$, $M_{200}$, and $N_G$, while the $1\sigma$ contours become narrower, as the FoF catalog contains several groups with low-to-intermediate $R_{200}$ and $M_{200}$ values that exhibit relatively low compactness (high $\zeta$), whereas the GraphShed groups generally show moderate-to-high compactness across the full range of $R_{200}$, $M_{200}$, and $N_G$. In addition, the KS tests applied to the $\zeta$ distributions across the different richness bins yield the $p$-values and $D$-statistics presented in the fourth row of Table \ref{tab:1}. Similar to the case of sphericity discussed earlier, all four richness bins exhibit statistically significant differences between the compactness distributions of the GraphShed and FoF groups, while the difference magnitude becomes larger from lowest to highest $N_{G}$ bins.

The comparison continues with the spin parameter of the groups. As discussed in previous studies (e.g., \citealp{hahn2007properties}), the spin parameter of a galaxy group is a dimensionless quantity originally introduced by \citet{peebles1969origin} and later revised for improved computational applicability by \citet{bullock2001universal}. This parameter characterizes the amount of ordered rotational motion relative to the internal random motions of the system and is defined as:
\begin{equation}
\label{Eq:5}
\lambda^{\prime} \equiv \frac{|\mathbf{J}_{200}|}{\sqrt{2GR_{200}}M_{200}^{3/2}}
\end{equation}
where $G$ denotes the Newtonian gravitational constant, and $\mathbf{J}_{200}$ represents the angular momentum of the group members located within the $R_{200}$ radius. Higher values of $\lambda^{\prime}$, generally indicate systems possessing greater specific angular momentum which can be associated with recent mergers or anisotropic accretion (e.g., \citealp{hetznecker2006evolution}), whereas lower values are more typical of dynamically settled or rotationally stable configurations (e.g., \citealp{bett2007spin}). The distribution of the spin parameter for the group catalogs of GraphShed and FoF is shown in the third row of Figure \ref{fig_3} as a function of $R_{200}$ and $M_{200}$, with the medians and $1\sigma$ contours for the different $N_G$ bins indicated by different colors and line-styles. As illustrated in these panels, the two catalogs exhibit contrasting trends in the spin parameter with increasing $R_{200}$, $M_{200}$, and $N_G$. In general, systems with larger $R_{200}$, higher $M_{200}$, and greater richness, tend to exhibit lower spin values in the GraphShed catalog, whereas the FoF catalog shows the opposite trend. This behavior is primarily associated with a subset of intermediate-to-large $R_{200}$, $M_{200}$, and $N_G$ systems in the FoF catalog that exhibit relatively high spin values, whereas such systems generally show lower spin parameters in the GraphShed catalog. Consequently, the difference between the two catalogs becomes more pronounced toward larger and richer systems, with the most substantial separation evident at high $R_{200}$, $M_{200}$, and $N_G$. The corresponding KS tests applied to the $\lambda^{\prime}$ distributions across the four richness bins yield the $p$-values and $D$-statistics presented in the fifth row of Table \ref{tab:1}. A statistically significant difference between the two spin-parameter distributions is found in the lowest, third, and highest richness bins, whereas the $10 < N_G \leq 20$ bin does not show a statistically significant difference. Consistent with the results for sphericity and compactness, the $D$-statistic further indicates that the magnitude of the distributional difference is larger in the higher-richness bins. This result is consistent with the distinct trends of the two catalogs toward larger and richer systems shown in Figure \ref{fig_3}.

The final property examined is the centroid shift $\Delta \mathcal{C}$ of the groups identified in the two catalogs. Although centroid shift has been defined and employed in different ways in previous studies (e.g., \citealp{poole2006impact,Ansarifard:2019ize}), in this study it is defined as the distance between the group center and the peak of the stellar mass density field:
\begin{equation}
\label{Eq:6}
\Delta \mathcal{C} \equiv ||\vec{r}_{\mathrm{MMD}} - \vec{r}_{\mathrm{UM}}||
\end{equation}
where, $\vec{r}_{\mathrm{MMD}}$ and $\vec{r}_{\mathrm{UM}}$ denote the position vectors of the center of the Voronoi cell with the Maximum Mass Density (MMD) within each group and the Unweighted Mean position (UM) of the member galaxies, respectively. These two positions are adopted to quantify the degree of asymmetry in the spatial distribution of stellar mass within the group. As mentioned in prior studies (e.g., \citealp{shin2021mass,korytov2023modeling}), the UM position can be used as a proxy for the halo center when satellite galaxies are not included. In the present study, low-mass and low-luminosity satellite galaxies are excluded from the sample (see Section \ref{sec:data}), thereby reducing potential biases arising from numerous faint satellites. In contrast, MMD position traces the location of the peak of the stellar mass density field within the group region. Considering Equation (\ref{Eq:6}), large values of $\Delta \mathcal{C}$ correspond to systems in which the stellar mass density peak is significantly displaced from the group center, indicating a more asymmetric mass distribution. Conversely, small values of $\Delta \mathcal{C}$ indicate systems in which the stellar mass peak and the group center are closely aligned, consistent with a more symmetric configuration. The distributions of $\Delta\mathcal{C}$ for the systems identified by GraphShed and FoF are shown in the bottom row of Figure \ref{fig_3} with respect to the $R_{200}$ and $M_{200}$, following the same configuration as for the previous parameters discussed above. The medians and $1\sigma$ contours for the different $N_G$ bins are depicted using distinct colors and line-styles. As seen in these panels, the groups identified by GraphShed generally exhibit smaller centroid shifts and span a narrower range of $\Delta\mathcal{C}$ than those in the FoF catalog. This difference is primarily associated with the presence of several FoF groups with relatively large centroid shifts that are not observed among the GraphShed groups. At small and intermediate $R_{200}$, $M_{200}$, and $N_G$, the two catalogs exhibit broadly similar behavior, with no pronounced variation in $\Delta\mathcal{C}$. Toward larger systems, however, both catalogs show an increasing trend in centroid shift, which is more pronounced in the highest-$N_G$ bin of the FoF catalog compared with GraphShed. The corresponding KS tests applied to the $\Delta\mathcal{C}$ distributions across the different $N_G$ bins yield the $p$-values and $D$-statistics presented in the bottom row of Table \ref{tab:1}. In the two lowest-richness bins, the $p$-values are larger than the adopted significance level of $\alpha=0.05$, indicating no statistically significant difference between the centroid-shift distributions of the two catalogs. In contrast, the two higher-richness bins exhibit statistically significant differences. Moreover, the increase in the $D$-statistic across the two highest-richness bins indicates a larger magnitude of the distributional difference at higher group richness. This result is consistent with the behavior of two catalogs toward larger and richer systems illustrated in Figure \ref{fig_3}.

Finally, it is worth noting that these structural metrics are not independent of one another, and the results presented here are altogether broadly consistent with those of Section \ref{sec:res_cc}: as GraphShed removes the low-density outskirts of FoF groups and segments the remaining structures, the resulting groups tend to have smaller $R_{200}$ radii and $M_{200}$ masses, be more spherical and compact, have lower spin, and show smaller offsets between their mass peaks and mean positions. These trends are consistent with the structural differences found here between the two catalogs. Furthermore, the results presented here indicate that the differences in the shape and dynamical properties of the groups generally become more pronounced toward larger $R_{200}$, $M_{200}$, and $N_G$, with the largest differences occurring for the most massive, extended, and richest systems (as depicted in Figure \ref{fig_3} and Table \ref{tab:1}). This behavior is also consistent with the results depicted in Section \ref{sec:res_cc}, where the differences between two group catalogs become more evident for larger systems.

\subsection{Large-Scale Statistics
\label{sec:LSS}}
Following the comparison of the physical and morphological properties of the groups identified by GraphShed and FoF, the next step examines how these group-finding methods affect the large-scale statistical properties derived from the resulting group catalogs. For this purpose, the un-weighted two-point correlation function (TPCF) and the mass function (MF) of the catalogs are compared (Table \ref{tab:0}).

\begin{figure}
	\centering
	\includegraphics[width=0.47\textwidth]{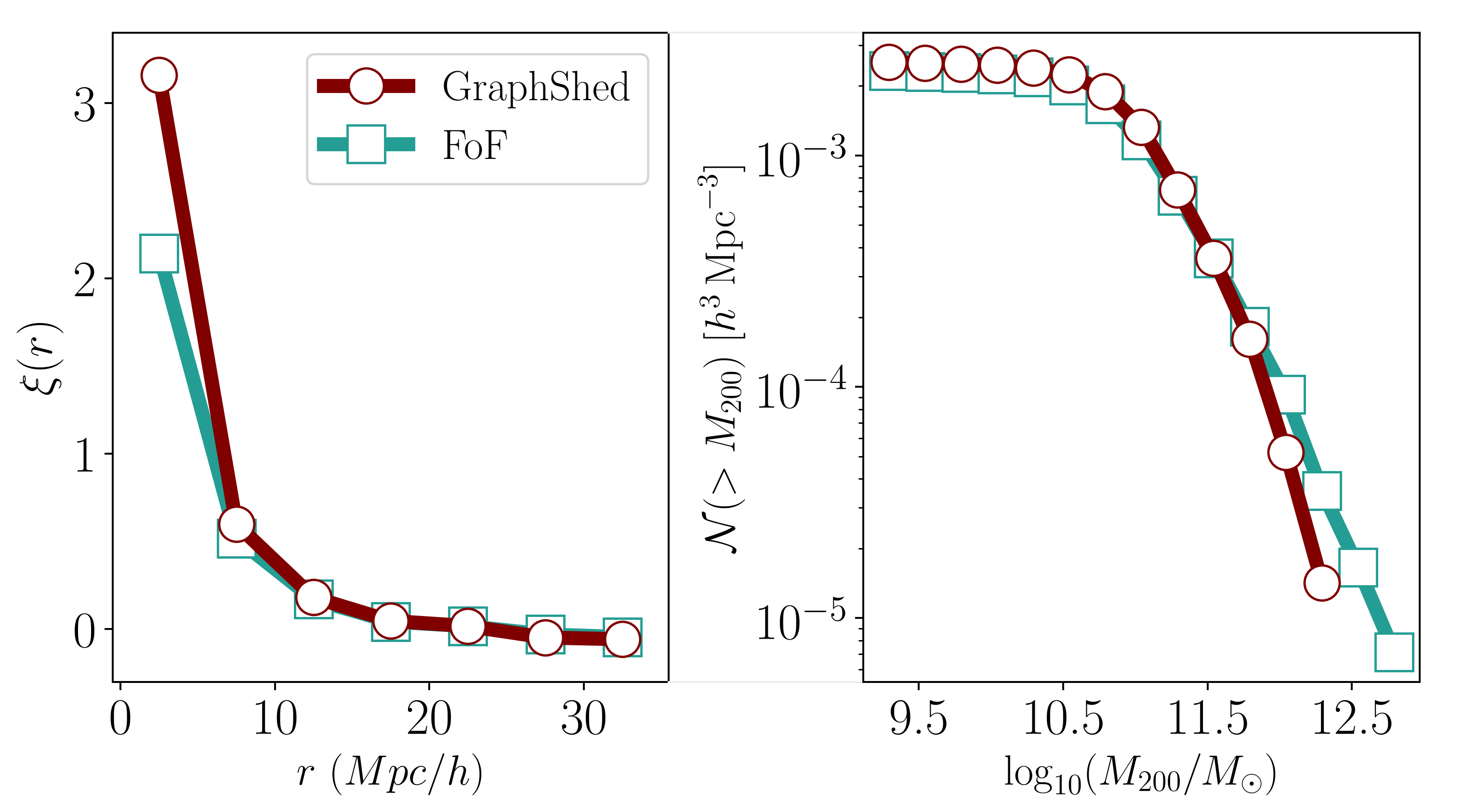}
	\caption{The two-point correlation function ($Left$ panel) as a function of separation distance ($r$) and the corresponding cumulative mass functions ($Right$ panel) as a function of stellar mass ($M$) for the group catalogs identified by the GraphShed and FoF methods are shown in dark red and light blue, respectively.}
	\label{fig_5}
\end{figure}

The TPCF, $\xi$, is calculated using the Landy-Szalay estimator (\citealp{landy1993bias}):
\begin{equation}
	\label{Eq:TPCF}
	\xi(r) = \frac{DD - 2DR + RR}{RR}
\end{equation}
where $D$ and $R$ denote the normalized pair counts at separation distance $r$, computed from the COM positions of the identified groups. Separations are measured up to half the simulation box length to minimize boundary effects. The resulting large-scale correlation functions are shown in the left panel of Figure \ref{fig_5}. In addition, the cumulative mass function (MF) of the groups $\mathcal{N}(>M_{200})$ is presented in the right panel of Figure \ref{fig_5} as a function of $M_{200}$, calculated as the cumulative number density of groups above a given $M_{200}$ threshold normalized by the total simulation volume. As shown in Figure \ref{fig_5}, the TPCFs of the GraphShed and FoF catalogs exhibit a similar behavior at large separation distances, while at smaller separations, the GraphShed catalog shows higher correlation amplitudes. This enhanced small-scale signal is most clearly evident in the first bin, with a weaker excess still apparent in the second bin, which originates from the density‑field–based nature of the GraphShed approach (as discussed earlier at Section \ref{sec:Grid_Vor}). Unlike the position‑based FoF method, GraphShed is more sensitive to variations in the underlying density field and therefore separates nearby over‑densities that FoF could merge into a single system (as mentioned at Section \ref{sec:res_cc} and Section \ref{sec:res_fof}).

To further investigate the difference between the TPCFs at small separations, particularly in the first two distance bins, a more detailed analysis is performed using the delete-one Jackknife method (\citealp{shao1986discussion}) to estimate the $\xi$ and its corresponding uncertainties. For this analysis, smaller separation bins are adopted, extending from $0.1$ to $10$ Mpc/ in $15$ logarithmic distance $r$ bins. The Jackknife realizations are constructed by dividing the total volume of the input dataset (Section \ref{sec:data}) into $N_{JK}$ equal-sized subvolumes and removing one subvolume at a time, while calculating the correlation function from the remaining data (e.g., \citealp{norberg2009statistical,mohammad2022creating}). This procedure produces $N_{JK}$ Jackknife realizations, each of which contains a fraction $(N_{JK}-1)/N_{JK}$ of the total volume. After calculating the TPCF for each Jackknife realization, the mean value of $\xi$ at each separation bin $r$ is given by:
\begin{equation}
	\label{Eq:xi_m}
	\bar{\xi}_{JK}(r) = \frac{1}{N_{JK}} \sum_{k=1}^{N_{JK}} \xi^{[k]}_{JK}(r).
\end{equation}
The covariance matrix between separation bins $r_i$ and $r_j$ is then estimated as:
\begin{equation}
	\label{Eq:xi_cov}
	\begin{aligned}
		C_{JK_{ij}} = \frac{N_{JK}-1}{N_{JK}} \sum_{k=1}^{N_{JK}} \left( \xi^{[k]}_{JK}(r_i)-\bar{\xi}_{JK}(r_i) \right) \\
		\qquad\times \left( \xi^{[k]}_{JK}(r_j)-\bar{\xi}_{JK}(r_j) \right)
	\end{aligned}
\end{equation}
where the factor $(N_{JK}-1)/N_{JK}$ accounts for the correlations between the Jackknife realizations arising from their substantial overlap, since each pair of realizations shares $N_{JK}-2$ of the $N_{JK}$ subvolumes (\citealp{tukey1958bias,miller1974jackknife}). Finally, the $1\sigma$ uncertainty in the TPCF $\xi$ at each separation bin $r$ is obtained by:
\begin{equation}
	\label{Eq:xi_err}
	\sigma_{JK}(r_i)=\sqrt{C_{JK_{ii}}}.
\end{equation}

Regarding the number of Jackknife subsamples, $N_{\rm JK}$, the corresponding Jackknife length needs to be sufficiently large compared with the maximum separation scale considered in the TPCF to minimize boundary effects (e.g., \citealp{beutler20116df}). A more conservative criterion is to restrict the use of Jackknife estimates to scales smaller than approximately half the size of the Jackknife length (e.g., \citealp{hildebrandt2017kids}). In the present analysis, the first two bins of the TPCFs shown in the right panel of Figure \ref{fig_5} extend to $10$ Mp/h. Adopting the conservative half-length criterion therefore requires the linear size of each Jackknife length to be larger than $20$ Mpc/h. Considering the $75$ Mpc/h side length of the IllustrisTNG100-1 simulation box (Section \ref{sec:data}), and noting that the Jackknife regions are obtained by dividing each axis into $n$ equal parts, such that $N_{JK}=n^3$, $n=2$ and $n=3$ are considered, corresponding to $8$ and $27$ Jackknife subsamples, respectively. These choices give Jackknife lengths of $37.5$ and $25$ Mpc/h, respectively, both of which satisfy the adopted scale requirement, whereas higher $n$ values would yield regions with linear lengths below this limit.

\begin{figure}
	\centering
	\includegraphics[width=0.4\textwidth]{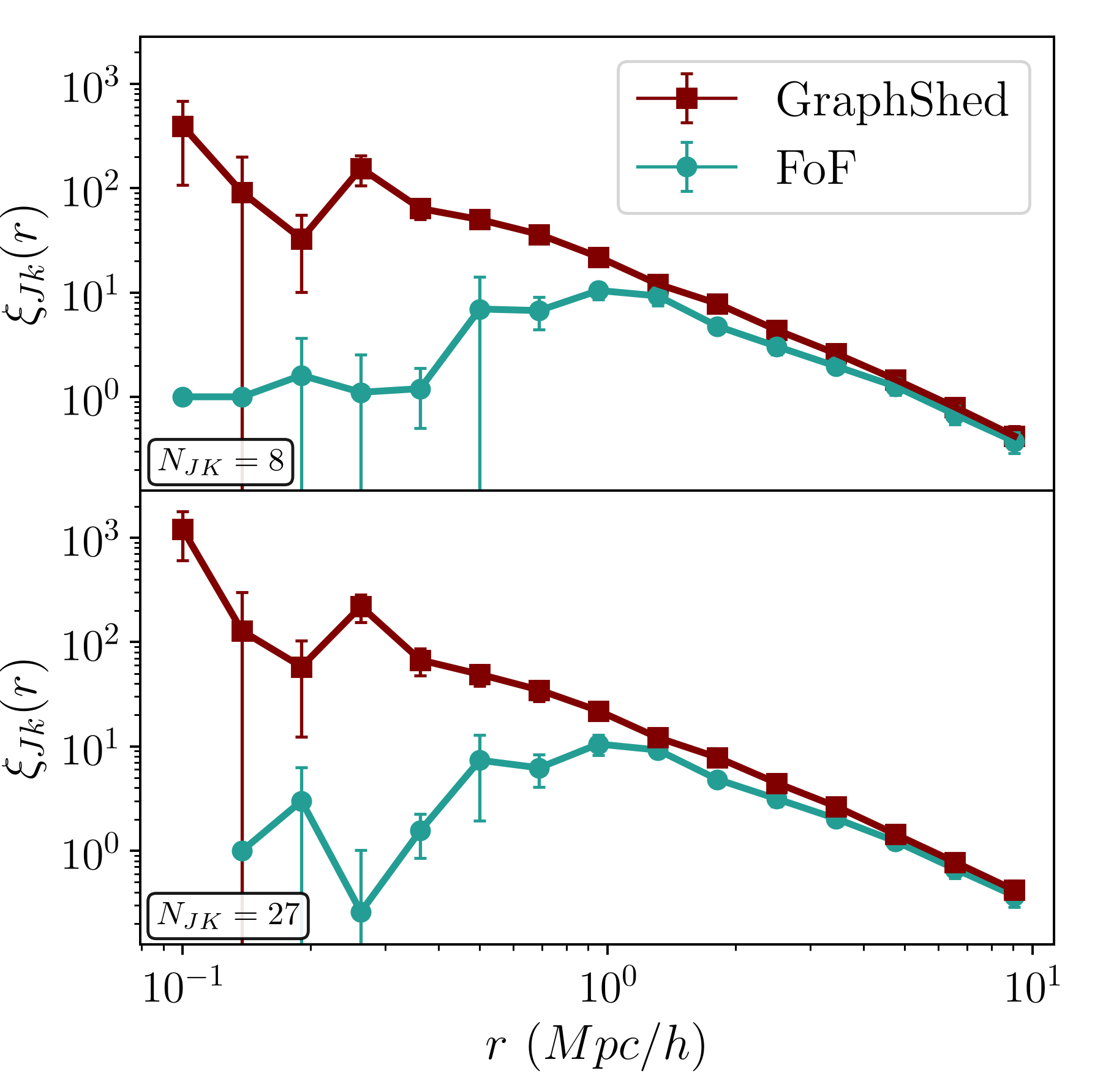}
	\caption{The Jackknife TPCFs $\xi_{JK}$ with the corresponding $1\sigma$ uncertainties as a function of the separation $r$ between the groups’ centers of mass. The GraphShed and FoF group catalogs presented in dark red and light blue, respectively. The $upper$ and $bottom$ panels correspond to $8$ and $27$ Jackknife subsamples.}
	\label{fig_xi}
\end{figure}

The resulting TPCFs and their associated uncertainties for $N_{JK}=8$ and $27$ are shown in the panels of Figure \ref{fig_xi}. As is evident from this figure, the GraphShed catalog exhibits higher correlation amplitudes than FoF over separations from $0.1$ to $\sim5$, consistent with the differences in shown at the left panel of Figure \ref{fig_5}. The most pronounced differences between the two catalogs occur at small separations, particularly in the range $0.1$ to $1$ Mpc/h. Within this range, the two catalogs exhibit contrasting trends: the correlation amplitude of the GraphShed groups generally increases toward smaller separations, whereas that of the FoF groups increases with separation up to $\sim1$ Mpc/h, after which it declines. This behavior can be attributed to the density-field-based nature of GraphShed, which segments FoF systems and distinguishes between nearby overdense aggregations that FoF instead groups together based on their spatial proximity. The upper and middle-left panels of Figure \ref{fig_6} illustrates an example of this excess of COM pairs at small separations for GraphShed groups relative to a FoF system.

In the case of the MF, the two catalogs show similar trends at low and intermediate $M_{200}$ values. However, for high-mass systems with $M_{200} \gtrsim$ $12$ $\mathrm{log}_{10}(M_{\odot}/h)$, the FoF catalog contains higher number of groups than the GraphShed catalog. Consistent with the results discussed previously, the FoF catalog also includes some systems with high $M_{200}$ values that are not present in the GraphShed catalog (as discussed at Section \ref{sec:res_fof}).

The similarities between the TPCFs and MFs obtained from the GraphShed and FoF catalogs indicate that the large‑scale statistical properties of the identified groups remain largely consistent between the two methods. Although the differences appear at small separations in the correlation function and at the high‑mass end of the mass function, the overall trends are in good agreement. 

As discussed previously in Section \ref{sec:res_fof}, the $M_{200}$ mass distributions of the GraphShed and FoF catalogs exhibit a high degree of statistical similarity. Together with the strong similarity in the TPCF at large separations and MF presented in this section, this suggests that cosmological interpretations and inferences derived from galaxy group catalogs (e.g., \citealp{chiu2023cosmological,bocquet2024spt}) are unlikely to be significantly affected, and that the resulting cosmological conclusions remain largely consistent. The close agreement with the FoF results further supports the use of GraphShed for studies that employ galaxy group catalogs as probes for cosmological inferences (e.g., \citealp{wang2022halo,seppi2024srg}).

\subsection{Merger Identification\label{sec:merger}}
\begin{figure*}
	\centering
	\includegraphics[width=0.8\textwidth]{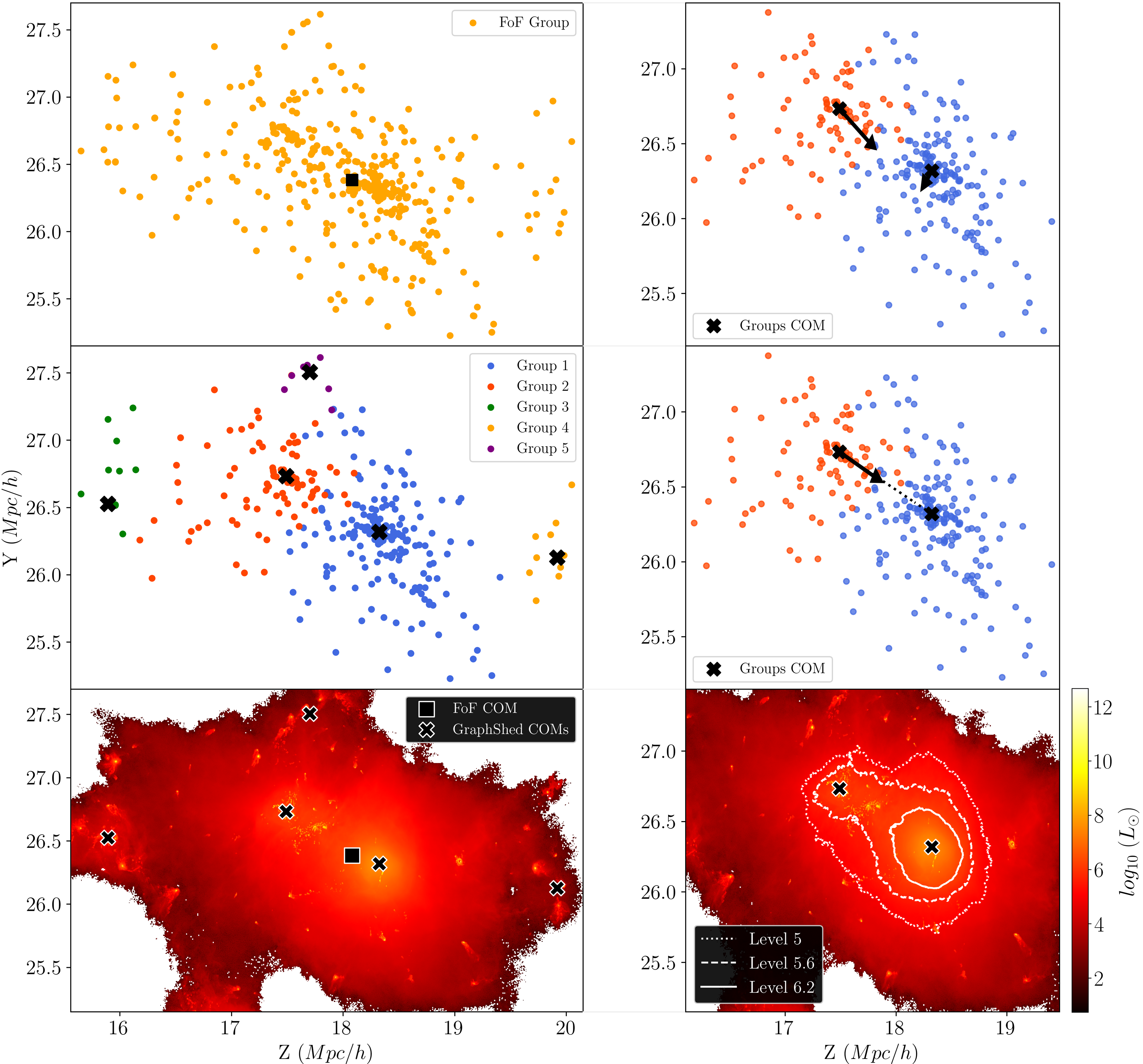}
	\caption{The stacked two-dimensional distribution of the galaxies in the largest FoF-identified group from the dataset of Section \ref{sec:data} is shown at the $Upper$-$Left$ panel, with the group’s center of mass marked by a black square. The projection stacks a $2.13$ Mpc/h along the $X$-direction, with the $Y$-$Z$ plane shown due to their larger extent. The GraphShed-detected groups at the same positional range are shown in the $Middle$-$Left$ panel. Each group plotted in a distinct color and their centers of mass indicated by black crosses. The X-ray luminosity map of the same region is presented in the $Bottom$-$Left$ panel, where the centers of mass of both the GraphShed groups and the FoF group are also marked with black crosses and a black square, respectively. The $Upper$-$Right$ panel shows the galaxies belonging to the two groups forming a merger-candidate pair, from among the GraphShed-identified groups and each shown in a different color. In addition, the weighted mean velocity vectors of the individual groups (Equation \ref{Eq:8}) are drawn as arrows originating from their respective centers of mass. The $Middle$-$Right$ panel shows the same galaxy distribution, but with the relative weighted velocity vector of the smaller group with respect to the larger system's rest frame (Equation (\ref{Eq:10})), drawn as an arrow originating from the smaller group's center of mass. The $Bottom$-$Right$ panel displays the X-ray luminosity map of the region containing two groups. The iso-contours are drawn at logarithmic levels of $5$, $5.6$ and $6.2$ in solar luminosity ($\mathrm{log}_{10}(L_{\odot})$), and for each level the largest contour is shown. The center of mass of the two groups are indicated with black crosses.\footnote{This map is derived from the IllustrisTNG100-1 simulation by computing the X-ray luminosity of the gas cells using the corresponding density maps and incorporating cooling tables presented by \cite{wiersma2009effect}.}}
	\label{fig_6}
\end{figure*}

The final aspect of the comparison between the GraphShed and FoF group catalogs is the identification and comparison of candidate mergers, flybys, and accretion events among the detected galaxy groups and clusters. For this purpose, a velocity‑based merger identification approach is introduced and employed. The proposed procedure is demonstrated using an example drawn from the group catalogs identified by the GraphShed (Section \ref{sec:GS}) and FoF methods, applied to the dataset described in Section \ref{sec:data} and illustrated in Figure \ref{fig_6}. As noted in Section \ref{sec:res_LL}, the corresponding linking lengths adopted by GraphShed and FoF for this dataset are very similar, ensuring that the presented comparison is not influenced by variations that could arise from differences in this parameter.

Among all pairs of galaxy groups and clusters in the resulting group catalogs identified by GraphShed and FoF, potential merger candidates are determined using the center‑of‑mass positions ($\vec{r}_{C}$) and the $R_{200}$ radii of the groups. A pair is classified as a potential merger candidate if the distance between the COMs of the two groups is smaller than or equal to the sum of their $R_{200}$ radii:
\begin{equation}
\label{Eq:7}
\Delta^{\alpha \beta} = |\vec{r}_{C}^{\ \alpha} - \vec{r}_{C}^{\ \beta}| \leq R_{200}^{\alpha} + R_{200}^{\beta} 
\end{equation}
where members of the pair are denoted by the labels $\alpha$ and $\beta$.

Then, for each potential merger candidate pair, the mass‑weighted radial and perpendicular relative velocities, $\Upsilon_{\parallel}$ and $\Upsilon_{\perp}$, are computed as described in Appendix \ref{App:Merg}. The pairs are subsequently classified according to these velocity components. A pair is flagged as a merger candidate if:
\begin{equation}
\label{Eq:16}
\Upsilon_{\parallel} > 0, \qquad |\Upsilon_{\parallel}| > |\Upsilon_{\perp}|
\end{equation}
indicating that the smaller group is approaching the larger system with a dominant radial motion. A receding pair ($\Upsilon_{\parallel} < 0$) with a dominant radial motion ($|\Upsilon_{\parallel}| > |\Upsilon_{\perp}|$) is therefore not considered a merger candidate. Conversely, the pair is classified as a flyby candidate if:
\begin{equation}
\label{Eq:17}
|\Upsilon_{\perp}| > |\Upsilon_{\parallel}|
\end{equation}
indicating a dominant perpendicular motion of the smaller system relative to the larger group. It should be noted that within this classification, some interacting systems such as pairs orbiting each other during the stages of a merger might also be categorized as flybys in the analyzed snapshot.

Finally, the mass ratio of each merger candidate pair (Equation (\ref{Eq:16})) is defined as (e.g., \citealp{contreras2022three}):
\begin{equation}
\label{Eq:18}
\Xi \equiv \frac{ \min \left( M_{200}^{\alpha}, M_{200}^{\beta} \right) }{ \max \left( M_{200}^{\alpha}, M_{200}^{\beta} \right) }
\end{equation}
and pairs are then classified according to their mass ratio. If $\Xi > 0.33$, the pair is considered as a major merger candidate, while those with $0.1 \leq \Xi \leq 0.33$ are considered as minor merger candidates (e.g., \citealp{contreras2022three}). At the end, pairs with $\Xi < 0.1$ are classified as accretion candidates, since the secondary component is sufficiently low in mass relative to the primary, that it does not constitute a genuine merger (e.g., \citealp{genel2010growth}).

After applying this procedure, a total of $271$ and $42$ potential group pairs satisfying the condition of Equation (\ref{Eq:7}) are detected and analyzed from the GraphShed and FoF group catalogs, respectively. The resulting classifications are summarized in Table \ref{tab:2}. Notably, none of the potential merger candidate pairs (Equation \ref{Eq:7}) were receding with a dominant radial component; all pairs in both catalogs are therefore classified into one of the mentioned categories. The substantially larger number of candidate pairs identified by GraphShed suggests the higher sensitivity of this density‑based approach to substructures within galaxy systems, enabling the detection of a broader range of dynamical interactions among nearby groups and clusters.

\begin{deluxetable}{c|c|c}  
	\tablewidth{0pt}
	\tablecaption{Classifications of galaxy group merger and interaction types for catalogs generated by GraphShed and FoF methods. \label{tab:2}}
	\tablehead{
		\multicolumn{1}{c}{$ $} & \multicolumn{2}{c}{N}\\
		\multicolumn{1}{c}{Class} & \multicolumn{1}{c}{GraphShed (271)} & \multicolumn{1}{c}{FoF (42)}}  
	\startdata
	Major mergers & $49$ & $8$ \\
	Minor mergers & $31$ & $13$ \\
	Accretions & $29$ & $9$ \\
	Flybys & $162$ & $12$ \\
	\enddata
\end{deluxetable}

A comparison of the merger categories shows that both methods detect comparable fractions of major mergers, with GraphShed and FoF revealing $49$ ($18\%$) and $8$ ($19\%$) pairs, respectively. In contrast, notable differences appear in the fractions of minor mergers and accretion events. The FoF catalog shows relatively larger proportions in these categories, accounting for $31\%$ and $21\%$ of the identified pairs respectively, whereas the GraphShed catalog yields smaller fractions of $\sim 11\%$ for both categories.

The most prominent difference between the two catalogs appears in the classification of flyby systems. In the GraphShed catalog, flybys constitute the majority of the detected pairs ($60\%$), while in the FoF catalog they represent about $29\%$ of the total. This indicates that GraphShed resolves a larger number of dynamically independent neighboring structures whose relative motions are dominated by tangential components rather than radial. The lower fraction of flybys identified in the FoF catalog reflects the tendency of this method to link nearby systems based solely on their spatial proximity, potentially combining galaxy systems that are in fact passing alongside each other (e.g., \citealp{kang2005semianalytical,bett2007spin}), consistent with the results discussed at previous subsections.

These results suggest that the higher sensitivity of density-based group-finding approaches can reveal not only merger candidates that remain hidden in position-only methods, but also identify flyby and accretion systems that might otherwise be grouped with other galaxy systems due to their positional proximity. Consequently, the GraphShed method provides a more detailed view of the dynamical interactions occurring within dense environments, enabling a clearer distinction between different types of close encounters among galaxy groups and clusters.

An example illustrating this capability is shown in Figure \ref{fig_6}. In this figure, the largest group identified by the FoF method from the galaxy distribution described in Section \ref{sec:data} is displayed. For comparison, the galaxy groups identified using the GraphShed approach, whose boundaries lie within the coordinate range of the FoF-identified group, are also shown. In addition, the cumulative X-ray luminosity map of this region is presented.

As shown in the left column of Figure \ref{fig_6}, the large group identified by the FoF method as a single irregular structure is separated into five distinct groups by the GraphShed algorithm. These groups correspond to different overdense regions that are located close to each other within that same coordinate range.  The presence of these distinct over-densities can also be inferred from the X-ray luminosity map, which could indicate that each group corresponds to a separate mass concentration, and the GraphShed method properly distinguishes these structures despite their spatial proximity.

Among these five GraphShed-identified groups, the pair flagged as a merger candidate by the velocity-based merger identification procedure described earlier is illustrated in the right column of Figure \ref{fig_6}. The X-ray luminosity map of this region, derived from the emission of gas cells in the simulation, together with fixed-level luminosity contours shown at several intensity levels, provides additional insights about these two systems. At the lowest contour level ($10^{5} L_{\odot}$), the emission exhibits an asymmetric, extended and bimodal morphology that encompasses the central region of both systems, suggesting the presence of multiple X-ray peaks associated with distinct over-densities (\citealp{jones1999einstein}). At the intermediate level ($10^{5.6} L_{\odot}$), the contour reveals a clear dumbbell-like structure with two connected lobes, indicating two nearby halos whose gaseous envelopes partially overlap. Such bimodal X-ray morphologies are commonly associated with interacting systems, containing more than one gravitational potential well (\citealp{jones1992clusters}). Finally, the highest contour level ($10^{6.2} L_{\odot}$) encloses only the central region of the brighter system, reflecting the deeper gravitational potential of the more massive group. The presence of two nearby X-ray peaks together with the elongated emission and the bridge of gas between them supports the interpretation of a merging pair (\citealp{andrade2015chandra}), consistent with the merger candidate identified by the velocity-based procedure described earlier.

\section{Summary and Conclusion\label{sec:sum}}
In this study, the GraphShed group finder is introduced as a data-derived, Voronoi-induced graph–based watershed approach designed to identify particle aggregations within the input dataset directly from its density field. The method has several key characteristics:
\begin{itemize}
	\item Employs the Voronoi tessellation on the combined set of grid points and data points to enhance the segmentation of the dataset’s spatial domain.
	\item Constructs a set of separated graphs using the adjacency of the Voronoi cells associated with the data points.
	\item Applies a top–down watershed procedure to the density field of the dataset in order to identify the particle aggregations.
	\item Computes all relevant quantities directly from the data, enabling the identification of groups without externally imposed thresholds or tunable parameters.
	\item Possesses a mathematically simple and computationally efficient formulation that is straightforward to implement.
\end{itemize}

A galaxy group catalog constructed by applying GraphShed to this dataset is systematically compared with the catalog obtained using the Friends‑of‑Friends (FoF) method on the same data (Section \ref{sec:res}). The comparison focuses on five main aspects: the difference between the linking lengths of the two methods across various sample selections, the cross-catalog agreement of galaxy memberships, the internal structural properties of the identified groups, the large-scale statistical measures of the galaxy groups and clusters (see, Table \ref{tab:0}), and the dynamical classification of interacting systems, including mergers, accretion events, and flybys.

The GraphShed-derived linking length is compared to that of FoF using multiple r-band absolute magnitude and stellar mass cuts (Section \ref{sec:res_LL}). These samples exhibit different global clustering powers, and the results show that the GraphShed-derived linking length adapts to the sample's clustering power while remaining close to the FoF linking length. However, it varies by approximately $18 \%$ relative to FoF between the highest and lowest clustered samples analyzed, with lower values corresponding to higher clustered samples and vice versa.

The correspondence between galaxy memberships in the GraphShed and FoF systems is quantified using a set of cross-catalog statistics, including purity, completeness, and removed fraction (Section \ref{sec:res_cc}). This comparison reveals that GraphShed finds $\sim 58 \%$ of the FoF systems identically, completely removes about $5 \%$ of them, and for the remainder, it shaves the low-density outskirts off the FoF systems and segments them if they contain multiple nearby over-densities. In addition, the dependence of these metrics on group mass is also evident, indicating that the strongest discrepancies between the two catalogs are preferentially associated with massive systems.

Several structural features are examined (Section \ref{sec:res_fof}), including the distributions of $R_{200}$, $M_{200}$, sphericity ($\Theta$), compactness ($\zeta$), spin parameter ($\lambda^{\prime}$), and centroid shift ($\Delta \mathcal{C}$), along with two-sided KS tests applied to these distributions across different group-richness $N_{G}$ bins. The $M_{200}$ distributions of the GraphShed and FoF groups show no statistically significant differences across any richness bin, while the remaining structural properties exhibit statistically significant differences in at least some $N_{G}$ bins. In particular, the differences in sphericity and compactness are significant across all $N_{G}$ bins, whereas the differences in, spin parameter, and centroid shift depend on group richness and generally become more pronounced toward larger and richer systems.

Large-scale properties of the resulting group catalogs are further examined (Section \ref{sec:LSS}) using the TPCF and the MF. The TPCFs obtained from the GraphShed and FoF catalogs show differences at small separations but display similar trends at larger scales. A more detailed analysis of the small-scale TPCFs using the delete-one Jackknife method revealed a higher correlation amplitude for the GraphShed groups than for FoF, as well as opposite trends between the two catalogs, particularly at separations below $\sim 1$ Mpc/h. The mass functions also exhibit close overall agreement, with discrepancies appearing primarily at the high-mass end.

Differences between the two catalogs become more evident when the dynamical interactions within the systems identified in each catalog are examined. To interpret these interactions, a velocity‑based classification scheme is introduced and applied to the GraphShed and FoF group catalogs (Section \ref{sec:merger}; Appendix \ref{App:Merg}) to identify mergers, accretion events, and flybys based on the radial and perpendicular components of the mass‑weighted relative velocity of potential merger candidates (Equation (\ref{Eq:7})). When applied to both catalogs, the classification reveals that GraphShed identifies a substantially larger number of interacting pairs than FoF. This difference is consistent with the interpretation that FoF might combine nearby systems undergoing flyby interactions into a single system due to positional proximity, while GraphShed more frequently resolves them as separate interacting structures.

Overall, the results demonstrate that GraphShed provides a robust alternative to traditional group finding approaches. The presented analysis indicates that the GraphShed group finder:
\begin{itemize}
	\item Minimizes parameter–driven uncertainties that could affect traditional group–finding approaches by avoiding tunable linking lengths or density thresholds (Section \ref{sec:GS}).
	\item Features computational simplicity, making it well suited for applications to large cosmological datasets (Section \ref{sec:GS}).
	\item Adapts its linking length based on the global clustering power of the input sample (Section \ref{sec:res_LL}).
	\item Alters several internal structural characteristics of the identified systems while preserving a mass distribution broadly consistent with that obtained using the FoF method (Section \ref{sec:res_fof}).
	\item Introduces differences in the non‑linear regime and at the high‑mass end of the TPCF and MF of the resulting catalog, while showing similar trends to FoF in the large‑scales and linear regime (Section \ref{sec:LSS}).
	\item Identifies a significantly larger number of interacting systems, revealing higher numbers of major and minor mergers as well as accretions and flybys due to its high sensitivity to variations in the density field, which remains hidden in position‑only approaches (Section \ref{sec:merger})
\end{itemize}

Future applications of GraphShed to observational datasets (e.g., \citealp{desi2024early}), larger cosmological simulations (e.g., \citealp{pakmor2023millenniumtng,dolag2025encyclopedia}), and controlled simulations with analytically specified halo profiles and substructure (e.g., \citealp{knebe2011haloes}) would provide a more direct assessment of its ability to recover known halo properties and resolve substructures independently of the assumptions of a particular halo finder. Applying GraphShed to substantially larger datasets (e.g., \citealp{villaescusa2020quijote}), would allow its computational scaling and resource requirements to be assessed on modern high-performance computing systems and compared with those of other group-finding approaches. Further studies of the sensitivity of GraphShed to substructure, together with comparisons with other physical density-based approaches (e.g., \citealp{hahn2007properties,pomarede2017cosmic}), will help quantify the extent to which its density-field-based segmentation provides enhanced sensitivity to embedded and nearby structures compared with traditional position-based methods. These investigations will enable a more comprehensive assessment of GraphShed's capability to identify galaxy systems, characterize group morphology, probe environmental drivers of galaxy evolution, and improve cosmological inferences drawn from studies of the cosmic web.

\begin{acknowledgments}
The authors appreciate the anonymous reviewer who helped to improve this paper. We are also very grateful to M. Alakhras for his invaluable comments on different parts of this paper. In addition, we gratefully acknowledge the use of the IllustrisTNG100-1 simulation database and its associated data products, which constitute a valuable resource for investigations of large-scale structure and cosmic-web environments. The GraphShed group-finder introduced in this work will be made publicly available in future publications; in the meantime, it can be obtained from the authors upon reasonable request.
\end{acknowledgments}

\appendix
\section{Regular Cubic Steiner Points for Voronoi Tessellation Construction \label{App:Grid}}
In computational geometry, auxiliary points can be introduced into an existing point set to control the geometry of the resulting tessellation or mesh (e.g., \citealp{cheng2013delaunay}). Such points are commonly referred to as Steiner points in the context of Delaunay-based mesh generation and related geometric constructions (e.g., \citealp{erten2009quality}). Different regular arrangements of auxiliary points can be considered, including simple cubic, hexagonal, and other regular arrangements (e.g., \citealp{toth2017handbook}). GraphShed adopts a simple cubic grid because of its straightforward construction and uniform spacing in the three spatial directions.

Considering the original galaxy catalog point set as:
\begin{equation}
	\label{G:1}
	P_{\rm G}=\{\mathbf{x}_i\}_{i=1}^{N_{G}}
\end{equation}
and the set of auxiliary Steiner points to be:
\begin{equation}
	\label{G:2}
	P_{\rm S}=\{\mathbf{s}_j\}_{j=1}^{N_{S}}
\end{equation}
with a uniform spacing $\ell$ between adjacent points (Equation \ref{Eq:1}), where the $N_{G}$ and $N_{S}$ are the total number of galaxies and auxiliary Steiner points, respectively. The Voronoi tessellation is then constructed from the combined point set:
\begin{equation}
	\label{G:3}
	P_{\rm T} =P_{\rm G}\cup P_{\rm S}
\end{equation}
where for a galaxy or grid point with position $\mathbf{x}_i$, its Voronoi cell is therefore defined by:
\begin{equation}
	\label{G:4}
	V_i=
	\left\{
	\mathbf{x}\in\mathbb{R}^3:
	\|\mathbf{x}-\mathbf{x}_i\|
	\leq
	\|\mathbf{x}-\mathbf{p}\|,
	\quad
	\forall\,\mathbf{p}\in P_{\rm T} \setminus \{\mathbf{x}_i\}
	\right\}.
\end{equation}

Therefore, these auxiliary grid points do not modify the Voronoi tessellation construction procedure; rather, they augment its input point set and consequently modify the geometry of the resulting cells.

This construction is particularly useful near the boundaries between overdense and underdense regions, where the absence of neighboring galaxies on the underdense side can otherwise lead to excessively large or irregular Voronoi cells (\citealp{ghafour2025vega}). The regular grid provides additional geometric sites that limit the spatial extent of such cells and produce a more controlled tessellation. The auxiliary points are massless and therefore do not contribute to the physical density assigned to the galaxy cells. Unlike adaptive Delaunay-refinement methods, the grid in GraphShed is prescribed a priori and is not iteratively optimized.

\section{Expected Mean Nearest-Neighbor Distance for $\delta_{CE}$
	\label{App:CE}}
The Clark-Evans index (\citealp{clark1954distance,clark1979generalization}) compares the observed mean nearest-neighbor distance of a point distribution to that expected under complete spatial randomness. While the original formulation was developed for 2D spatial point patterns, its extension to 3D is straightforward and is derived below.

A set of $N_{T}$ points uniformly and independently distributed within a total volume $V_{T}$ in $\mathbb{R}^3$ is considered. This corresponds to a 3D homogeneous Poisson process with intensity:
\begin{equation}
	\label{Eq:A1}
	\lambda = \frac{N_{T}}{V_{T}}.
\end{equation}

For a given point, the number of points within a sphere of radius $r$ follows a Poisson distribution with mean $\frac{4}{3}\pi \lambda r^{3}$. Therefore, the probability that there are no points within a distance $r$ is:
\begin{equation}
	\label{Eq:A2}
	P(\text{no points within } r) = e^{-\frac{4}{3}\pi \lambda r^3}.
\end{equation}

Let $R$ be the distance from a randomly chosen point to its nearest neighbor. The cumulative distribution function of $R$ is the probability that the nearest neighbor is at a distance less than or equal to $r$, which is equivalent to the probability that there is at least one point within the sphere of radius $r$:
\begin{equation}
	\label{Eq:A3}
	P(R \leq r) = 1 - P(\text{no points within } r) = 1 - e^{-\frac{4}{3}\pi \lambda r^3}.
\end{equation}
The probability density function (PDF) is obtained by differentiating the cumulative distribution function with respect to $r$:
\begin{equation}
	\label{Eq:A4}
	p(r) = \frac{d}{dr} P(R \leq r) = 4\pi \lambda r^2 e^{-\frac{4}{3}\pi \lambda r^3}.
\end{equation}
The expected (mean) nearest-neighbor distance is the first moment of the PDF:
\begin{equation}
	\label{Eq:A5}
	\langle d_{P}^{NN} \rangle = \int_0^\infty dr p(r) r.
\end{equation}
Substituting the PDF from Equation (\ref{Eq:A4}):
\begin{equation}
	\label{Eq:A7}
	\langle d_{P}^{NN} \rangle = 4\pi \lambda \int_0^\infty r^3 e^{-\frac{4}{3}\pi \lambda r^3} dr.
\end{equation}
Performing an integral by part, leads to:
\begin{equation}
	\label{Eq:A14}
	\langle d_{P}^{NN} \rangle = \Gamma\left(\frac{4}{3}\right) \left(\frac{3}{4\pi \lambda}\right)^{1/3}.
\end{equation}

Finally, substituting $\lambda$ from Equation (\ref{Eq:A1}):
\begin{equation}
	\label{Eq:A15}
	\langle d_{P}^{NN} \rangle = \Gamma\left(\frac{4}{3}\right) \left(\frac{3V_{T}}{4\pi N_{T}}\right)^{1/3}.
\end{equation}
This is the expected mean nearest-neighbor distance for a 3D homogeneous Poisson process (Equation (\ref{d_CE}) at Section \ref{sec:res_LL}), from which the global clustering power $\delta_{CE}$ (Equation (\ref{del_CE})) is defined.

\section{Mass-Weighted Radial And Perpendicular Relative Velocities 
	\label{App:Merg}}
The calculation of the mass-weighted radial and perpendicular components of the relative velocity proceeds in three steps. First, the velocity vectors of the galaxies in each group are used to compute the stellar‑mass–weighted mean velocity vector of that group:
\begin{equation}
\label{Eq:8}
\vec{V}_{\alpha} = \frac{\sum_{i \in \alpha} M_{i}^{*} \vec{v}_{i}}{\sum_{i \in \alpha} M_{i}^{*}},  \qquad
\vec{V}_{\beta} = \frac{\sum_{j \in \beta} M_{j}^{*} \vec{v}_{j}}{\sum_{j \in \beta} M_{j}^{*}}
\end{equation}
where $M^{*}$ and $\vec{v}$ denote the stellar mass and velocity vector of each member galaxy, respectively. The resulting weighted mean velocity vector of the groups are illustrated in the upper-right panel of Figure \ref{fig_6}, originating from the COM of each group.

In the second step, the larger system is identified based on the total stellar mass of the member galaxies within each system. The mean weighted velocity vectors of the groups are then transformed to the rest frame of the larger system. This is achieved by subtracting the weighted mean velocity vector of the larger group from that of the two groups. As a result, the relative weighted velocity vector of the larger system vanishes, and that of the smaller group with respect to the larger group is obtained as:
\begin{equation}
\label{Eq:10}
\vec{\mathcal{V}}_{\alpha} = \vec{V}_{\alpha} - \vec{V}_{\beta}
\end{equation}
where $\beta$ is assumed to be the larger system ($\sum_{j \in \beta} M_{j}^{*} > \sum_{i \in \alpha} M_{i}^{*}$), and thus $\vec{\mathcal{V}}_{\beta}=0$. This relative weighted velocity vector is shown in the middle-right panel of Figure \ref{fig_6}, drawn from the COM of the smaller group.

In the third step, the radial unit vector is defined from the COM of the smaller group ($\vec{r}_{C}^{\ \alpha}$) toward the COM of the larger system ($\vec{r}_{C}^{\ \beta}$):
\begin{equation}
\label{Eq:11}
\hat{\tau}_{\alpha} = \frac{\vec{r}_{C}^{\ \beta} - \vec{r}_{C}^{\ \alpha}}{|\vec{r}_{C}^{\ \beta} - \vec{r}_{C}^{\ \alpha}|}
\end{equation}
The radial component of the relative weighted velocity vector is then obtained as:
\begin{equation}
\label{Eq:12}
\Upsilon_{\parallel} \equiv \vec{\mathcal{V}}_{\alpha} \cdot \hat{\tau}_{\alpha}
\end{equation}
and the perpendicular component is given by:
\begin{equation}
\label{Eq:13}
\Upsilon_{\perp} \equiv |\vec{\mathcal{V}}_{\alpha} - \Upsilon_{\parallel}\hat{\tau}_{\alpha}|
\end{equation}
These components are used in the merger and flyby conditions (Equations (\ref{Eq:16}) and (\ref{Eq:17})).

\bibliography{GraphShed_Refs.bib}{}
\bibliographystyle{aasjournalv7}

\end{document}